\documentclass[%
 reprint,
 superscriptaddress,
 showpacs,preprintnumbers,
 amsmath,amssymb,
 aps,
pra,
 floatfix,
]{revtex4-1}

\usepackage[T1]{fontenc} 
\usepackage{mathptmx}

\usepackage{graphicx}
\usepackage{dcolumn}
\usepackage{bm}
\usepackage[colorlinks,breaklinks,
 citecolor=red,linkcolor=red,urlcolor=blue
]{hyperref}

\begin{document}
\bibliographystyle{apsrev4-1}

\title{Structures, Branching Ratios and Laser Cooling Scheme for $^{138}$BaF Molecule}%

\author{Tao Chen}%
 \email{chentao1990@outlook.com}
\author{Wenhao Bu}
\author{Bo Yan}%
 \email{yanbohang@zju.edu.cn}
\affiliation{%
 Department of Physics, Zhejiang University, Hangzhou, China, 310027
}%


\date{\today}

\begin{abstract}
For laser cooling considerations, we have theoretically investigated the electronic, rovibrational and hyperfine structures of BaF molecule. The highly diagonal Franck-Condon factors and the branching ratios for all possible transitions within the lowest-lying four electronic states have also been calculated. Meanwhile, the mixing between metastable $A^\prime{}^2\Delta$ and $A^2\Pi$ states and further the lifetime of the $\Delta$ state have been estimated since the loss procedure via $\Delta$ state would like fatally destroy the main quasi-cycling $\Sigma - \Pi$ transition for cooling and trapping. The resultant hyperfine splittings of each rovibrational states in $X^2\Sigma^+$ state provide benchmarks for sideband modulations of cooling and repumping lasers and remixing microwaves to address all necessary levels. The calculated Zeeman shift and $g$-factors for both $X$ and $A$ states serve as benchmarks for selections of the trapping laser polarizations. Our study paves the way for future laser cooling and magneto-optical trapping of the BaF molecule.
\end{abstract}
\maketitle


\section{\label{section1}Introduction}

Ultracold polar molecules\cite{Carr2009}, due to the tunable long-range dipole-dipole interactions\cite{Yan2013}, provide access to lots of new potential regimes like novel many-body physics\cite{Baranov2012}, ultracold chemistry\cite{Ospelkaus2010}, precision measurement\cite{Chin2009} and quantum computation and infomation processing\cite{DeMille2002,Andre2006}. However, one great challenge is to produce quantum molecular samples (such as KRb\cite{Ni2008,Moses2015}), that means, to achieve high phase space density in ultracold regime. Besides complicated indirect forming methods, direct Stark, Zeeman and optoelectric slowing and cooling\cite{Bochinski2003,Narevicius2008,Fulton2006,Liu2012} could only yield molecular samples at milliKelvin temperature, and even worse thing is that the phase space density could not be increased. Consequently, approaches of extending widely-used traditional laser cooling technique in atoms to polar molecules are under exploration\cite{DiRosa2004,Stuhl2008}, and fortunately, the laser cooling and magneto-optical trapping experiments have recently been realized for particular species of molecules,including SrF\cite{Shuman2010,Barry2014}, YO\cite{Hummon2013,Yeo2015}, and CaF\cite{Zhelyazkova2014,Hemmerling2016}. Now other ongoing candidates, like YbF\cite{Smallman2014}, MgF\cite{Xu2016}, BH\cite{Hendricks2014}, RaF \cite{Isaev2010}, TiF \cite{Hunter2012} and BaH \cite{Tarallo2016},  have attracted great interests as well.

For molecules, the much more complex internal rovibrational structures make finding a closed cycling transition required by laser cooling much more difficult than for atoms. Those laser-cooled molecules above share a common feather, highly diagonal Franck-Condon factors (FCFs), which results in a relatively simple quasi-cycling cooling scheme, that is, only two repmuping laser beams are required experimentally \cite{Barry2014} to eliminate undesired spontaneous decays to dark vibrational states. Besides the vibrational transitions, rotational transitions could be closed by choosing the $\vert N=1\rangle\leftrightarrow\vert N=0\rangle$ transition where the strict parity and angular momentum selection rules guarantees the pumped excited-state molecules entirely decay back to $N=1$ ground state \cite{Stuhl2008}. Meanwhile, all the hyperfine levels are addressed by sideband modulations of the pumping and repumping lasers\cite{Shuman2010}, and the problem induced by Zeeman dark state has been solved by either adding an angled static magnetic field \cite{Shuman2010} or using the so-called switching scheme\cite{Hummon2013,Norrgard2016}. Another point is the loss channel via the metastable $\Delta$ state to even parity $N=0,2$ rotational states, which must be remixed into the optical cycling with microwave dressing, as performed in YO experiments\cite{Yeo2015}.

\begin{figure}[]
\includegraphics[width= 0.45\textwidth]{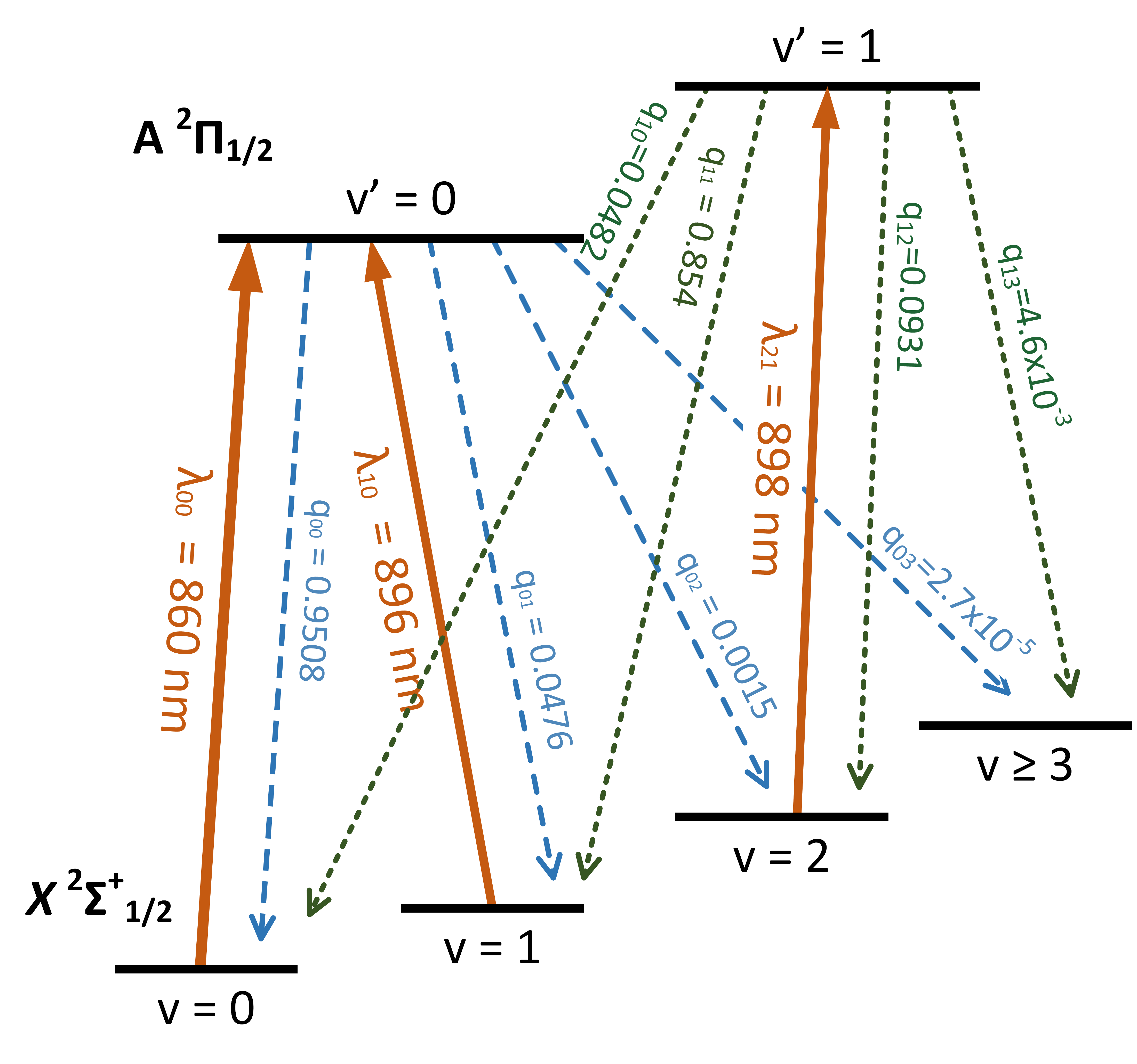}
\caption{\label{figure1}(Color online) Vibrational branching is suppressed to achieve a quasi-cycling transition for laser cooling BaF molecule. The orange lines indicate the transitions driven by the cooling laser $\lambda_{00}$ and the repumping lasers $\lambda_{10}$, $\lambda_{21}$. The blue and green lines indicate the spontaneous decays from $A(v'=0$ and $v'=1)$ to vibrational states in $X$, respectively. $q_{v'v}$ represents the FCFs for the transition $\vert A,v'\rangle \to \vert X,v\rangle$.}
\end{figure}

Extending the laser cooling technique to more polar molecules is a challenging and hot topic in physics. BaF molecule is a good and promising candidate. It can be used to study parity violation \cite{DeMille2008} . The involving transitions have the wavelengths around $900$ nm (see Fig.\ref{figure1}), in the good regime of the diode laser. It is easy to get high power laser with low cost, making the laser cooling experiments much simpler. Here we consider the feasibility for laser cooling and trapping of BaF molecule. Firstly, the electronic excited $A^2\Pi_{1/2}$ state has a short lifetime of $\tau\thicksim$ 56 ns (the natural linewidth $\Gamma\approx2\pi\times3$ MHz)\cite{Berg1998}, which enables large photon scattering rates. Meanwhile, our calculation on the FCFs of $X^2\Sigma_{1/2}^+\leftrightarrow A^2\Pi_{1/2}$ transition shows that BaF indeed possesses the common highly diagonal feather; see details in Sec.\ref{section2}. Since BaF molecule has metastable $\Delta$ state, for sideband modulation and microwave remixing considerations, we employ an effective Hamiltonian to obtain the energy values of the spin-rotation and hyperfine levels in $X^2\Sigma$ state, and further propose a sideband modulation and microwave addressing scheme in Sec.\ref{section3}. According to the magneto-optical trapping designs for SrF and YO\cite{Barry2014,Hummon2013}, we calculated the branching ratios from Zeeman sublevels in $A^2\Pi_{1/2}$ to those in $X$ state, g-factors for each hyperfine levels, and Zeeman splitting under external magnetic field; see Sec.\ref{section4} and \ref{section5} respectively. In Sec.\ref{section6}, the metastable $\Delta$ state has been detailly investigated due to the possible decay to other even parity rotational states. We mainly focus on the mixing with $A^2\Pi_{1/2,3/2}$ states and the branching ratios to vibrational states in $X$ ground state. Section \ref{section7} gives a brief conclusion to this work.

\section{\label{section2}Laser cooling scheme and Franck-Condon factors for \emph{A-X} transition}

The BaF molecule has similar electronic structures with SrF \cite{Shuman2010} and YO \cite{Bernard1992}, and so does the laser cooling scheme, as shown in Fig.\ref{figure1}. The quasi-cycling transition here temporarily takes no consideration on the $\Delta$ state which will be discussed in Sec.\ref{section6}. Suppression of vibrational branching requires highly diagonal FCFs for $A\rightarrow X$ transitions. We have performed a careful calculation of the FCFs: Firstly, we numerically modeled the potential energy curves of the lower-lying $\Sigma$, $\Delta$ and $\Pi$ states with the corresponding parameters as listed in Table \ref{table1} by employing the RKR (Rydberg-Klein-Rees) method\cite{Rees1947}. We checked the RKR potential curves with the analytical Morse potentials, they are almost the same, especially at the region near the equilibrium positions. Then, we use the Symplectic propagation method to solve the Sch\"odinger equation and meanwhile make an eigenenergy correction to the trial energy $E=\omega_e(v+1/2)-\omega_e\chi_e(v+1/2)^2$; see Ref.\cite{Chen2014} for more details. After getting the wavefunctions of each vibrational state, we finally calculate the overlap integrals $\tilde{q}_{v'v}=\langle v'\vert v\rangle$ and FCFs $q_{v'v}=\vert\langle v'\vert v\rangle\vert^2$, where $\vert v\rangle$ is the vibrational wavefunction.

The related values are shown in Fig.\ref{figure1}. Since the $\vert A, v^\prime=0\rangle\to\vert X, v\ge 3\rangle$ branching is $q_{03}=2.7\times10^{-5}$, we use $\vert X, v=0\rangle \to\vert A, v'=0\rangle$ transition as the main pumping, the $\vert X, v=1\rangle \to\vert A, v'=0\rangle$ and $\vert X, v=2\rangle \to\vert A, v'=1\rangle$ transitions as the first- and second-stage repumping to achieve $ \thicksim3\times10^4$ photons scattering before molecules populate $\vert X, v\ge3\rangle$ levels. The wavelengths are $\lambda_{00}=860 \text{nm}$, $\lambda_{10}=896 \text{nm}$ and $\lambda_{21}=898 \text{nm}$ respectively. The more accurate wavelength values are listed in Table \ref{table0}. Specially, the experimental value for $\lambda_{00}$ is derived from the measured spectroscopy data \cite{Steimle2011}. The difference between our calculation and the experimental value is as small as 0.038 cm$^{-1}$.  In future experiments, a repumping laser for $\vert X,v=3\rangle\to\vert A,v'=2\rangle$ might be preferred to enhance the photon scattering, just like the 3D magneto-optical trapping experiment of SrF \cite{Barry2014}.

\begin{table}[]
\caption{\label{table0}
Accurate calculated wavelength values for the cooling and repumping lasers in Fig.\ref{figure1}. The values are generated with the $T_e$, $A_e$ values in Table \ref{table1} and the calculated eigenenergy values for corresponding vibrational states. The experimental wavelength for $\lambda_{00}$ is derived from the measured spectroscopy data $R_1(0)$ for $\vert X,v=0,N=0,+\rangle\to\vert A,v=0,J=1/2,-\rangle$ transitions in Ref.\cite{Steimle2011} and the $B_e$ values in Table \ref{table1}. }
\begin{ruledtabular}
\begin{tabular}{ccc}
Transitions & calculated [nm] & from experiment [nm]\\\hline
$\lambda_{00}$ & 859.79289 & 859.79569 \\
$\lambda_{10}$ & 895.65807 & \\
$\lambda_{21}$ & 897.89917 & \\
$\lambda_{32}$ & 900.15402 & \\
\end{tabular}
\end{ruledtabular}
\end{table}

\begin{table}[]
\caption{\label{table1}
Parameters of the lower-lying electronic states of the $^{138}$BaF molecule from previous experimental data \cite{Bernard1992}. The $T_e$ and splitting constant $A_e$ result in consistent values of $T_e$ in Ref.\cite{Barrow1988}. The $\omega_e$ and $\omega_e\chi_e$ values for the $\Pi_{1/2,3/2}$ and $\Delta_{3/2,5/2}$ doublets have a little difference; see Ref.\cite{Barrow1988} for details. All values here are in unit of cm$^{-1}$.}
\begin{ruledtabular}
\begin{tabular}{llll}
 & $X^2\Sigma$ & $A^{\prime 2}\Delta$ & $A^2\Pi$ \\
\hline
$T_e$ & 0 & 10940.27 & 11962.174 \\
$A_e$ & & 206.171 & 632.409 \\
$\omega_e $ & 469.4161 & 437.41 & 437.899 \\
$\omega_e\chi_e$ & 1.83727 & 1.833 & 1.854  \\
$\alpha_e\times 10^3$ & 1.163575 & 1.2052 & 1.2563 \\
$B_e$ & 0.21652967 & 0.210082 & 0.212416 \\
\end{tabular}
\end{ruledtabular}
\end{table}

Table \ref{table2} lists the FCFs $q_{v'v}$ for vibrational transitions from $A^2\Pi_{1/2}$ and $A^2\Pi_{3/2}$ to $X^2\Sigma_{1/2}$ respectively. Since the parameters $\omega_e$ and $\omega_e\chi_e$ for $A^2\Pi_{1/2}$ and $A^2\Pi_{3/2}$ states are nearly the same \cite{Barrow1988}, the corresponding values of the calculated FCFs have no significant differences with each other. Since the angular momentum selection rules forbid electronic dipole transitions in $A'^2\Delta\to X^2\Sigma_{1/2}$, the $\Delta$ state decay back to $X$ state via its mixing with the $\Pi$ states, and thus we do not list the FCFs here. The branching ratios for the spontaneous emissions from $A'$ to $X$ will be discussed in Sec.\ref{section6}.

\begin{table*}[t]
\caption{\label{table2} The calculated FCFs $(q_{v'v})$ for vibrational transitions in $\vert A^2\Pi_{1/2},v'\rangle\to\vert X^2\Sigma_{1/2},v\rangle$ and $\vert A^2\Pi_{3/2},v'\rangle\to\vert X^2\Sigma_{1/2},v\rangle$ using the RKR potentials.
}
\begin{ruledtabular}
\begin{tabular}{cccccc}
$ A^2\Pi_{1/2},v'\to X,v$ & $v'=0$ & $v'=1$ & $v'=2$ & $v'=3$ & $v'=4$ \\\hline
$v=0$ & 0.9508 & 0.0483 & $9.1\times10^{-4}$ & $1.9\times10^{-6}$ & $4.5\times10^{-7}$ \\
$v=1$ & 0.0476 & 0.8539 & 0.0956 & 0.0030 & $1.3\times10^{-5}$ \\
$v=2$ & $1.5\times10^{-3}$ & 0.0925 & 0.7581 & 0.1412 & 0.0065 \\
$v=3$ & $2.7\times10^{-5}$ & $5.1\times10^{-3}$ & 0.1347 & 0.6643 & 0.1841 \\
$v=4$ & $4.6\times10^{-7}$ & $1.3\times10^{-4}$ & 0.0104 & 0.1733 & 0.5738 \\
\hline\hline
$ A^2\Pi_{3/2},v'\to X,v$ & $v'=0$ & $v'=1$ & $v'=2$ & $v'=3$ & $v'=4$ \\\hline
$v=0$ & 0.9508 & 0.0482 & $9.7\times10^{-4}$ & $3.8\times10^{-6}$ & $2.1\times10^{-7}$ \\
$v=1$ & 0.0476 & 0.8539 & 0.0956 & 0.0032 & $2.2\times10^{-5}$ \\
$v=2$ & $1.6\times10^{-3}$ & 0.0928 & 0.7582 & 0.1404 & 0.0068 \\
$v=3$ & $2.9\times10^{-5}$ & $4.8\times10^{-3}$ & 0.1352 & 0.6648 & 0.1828 \\
$v=4$ & $3.7\times10^{-7}$ & $1.2\times10^{-4}$ & 0.0099 & 0.1743 & 0.5746
\end{tabular}
\end{ruledtabular}
\end{table*}

\section{\label{section3}Hyperfine structures}

\begin{figure}[b]
\includegraphics[width= 0.48\textwidth]{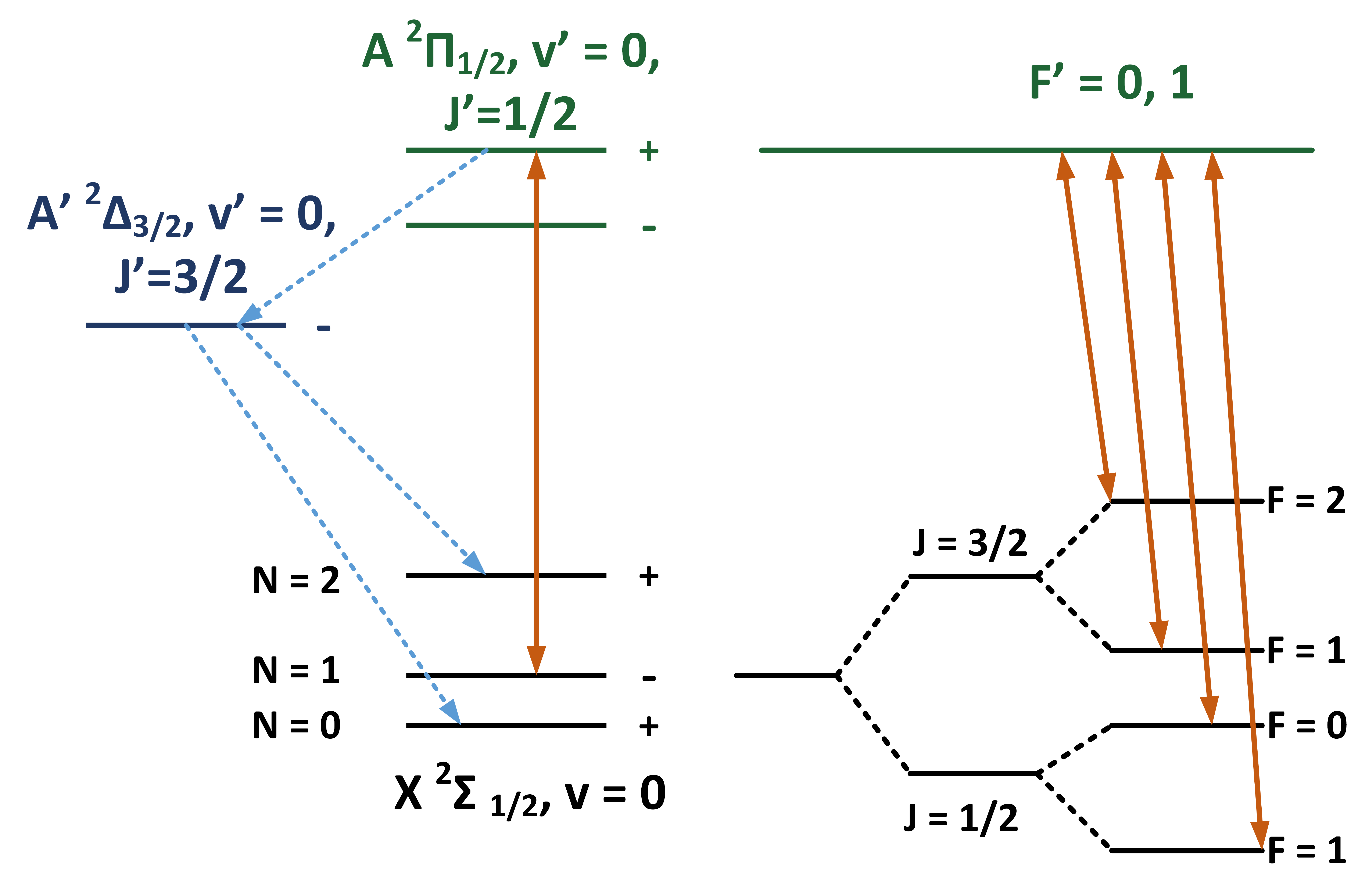}
\caption{\label{figure2}(Color online) Rotational branching is elimilated by driving $\vert X^2\Sigma_{1/2}, N=1, -\rangle \to \vert A^2\Pi_{1/2}, J'=1/2, +\rangle$ transition. Unfortunately, $\vert A^2\Pi_{1/2}, J'=1/2, +\rangle$ can also decay to $\vert A'^2\Delta_{3/2}, J'=3/2, -\rangle$ state, and then back to $\vert X, N=0,2, +\rangle$ rotational states, which will break the optical cycling. In addition, $\vert X, N=1, -\rangle$ state obeys Hund's case (b) and suffers from spin-rotaion and hyperfine splitting, thus the optical cycling should adress all possible levels. Here +(-) indicates even(odd) parity.}
\end{figure}

Now let us consider the rotational branchings. The $X^2\Sigma_{1/2}$ state is a Hund's case (b) state and rotational quantum number $N$ is a good quantum number, but for Hund's case (a) state $A^2\Pi_{1/2}$, the good quantum number is $J=N+\Omega$ coupled by rotational and electronic angular momentum. $\vert A^2\Pi_{1/2},J'=1/2\rangle$ state has a double orbital degeneracy \cite{Mulliken1931}, that is, $\Lambda$-doubling structures corresponding to odd- and even-parity electronic wavefunctions respectively. Following the parity and angular momentum selection rules for dipole transition, the parities of the initial and final states should be opposite and $\Delta J=0,\pm1$. Consequently, driving $\vert X, N=1, -\rangle \rightarrow \vert A, J'=1/2, +\rangle$ transitions results in only decays back to $N=1$ are allowed as shown in Fig.\ref{figure2}.

However, the hyperfine structures of both the ground $X$ and excited $A$ states should be taken into account for optical pumping, otherwise the dark states exist. For Hund's case (a) state $A^2\Pi_{1/2}$, the $\Lambda$-splitting for $J$ state with different parity is $\delta E_\Lambda=-(p+2q)(J+1/2)$, where $p+2q=-0.25755~\text{cm}^{-1}$ for $^{138}$BaF \cite{Steimle2011}; while the hyperfine splitting between $F'=0$ and $F'=1$ for $\vert J'=1/2,+\rangle$ is unresolved yet. For ground state $X^2\Sigma^+_{1/2}$, both the spin-rotation and hyperfine interactions split the $\vert X, N=1, -\rangle$ state into four components as shown in Fig.\ref{figure2}. All four hyperfine levels should be pumped simultaneously to prevent molecules accumulate into one state. To implement the sideband modulation, we have to calculate the relatively precise energy splittings of the hyperfine states in each rotational state of $X^2\Sigma_{1/2}$. The effective Hamiltonian contains the molecular rotational term $H_\text{R}$, spin-rotational coupling $H_\text{SR}$ and hyperfine interaction $H_\text{hfs}$, and is given by
\begin{eqnarray}
\begin{array}{lcl}
H_\text{eff}&=&H_\text{R}+H_\text{SR}+H_\text{hfs}, \\[0.5em]
H_\text{R}&=&B_v\hat{N}^2-D_{v}\hat{N}^4,\\[0.5em]
H_\text{SR}&=&\gamma_{v N}T^1(\hat{S})\cdot T^1(\hat{N}) ,\\[0.5em]
H_\text{hfs}&=&b_FT^1(\hat{I})\cdot T^1(\hat{S})
+c_vT^1_{q=0}(\hat{I})T^1_{q=0}(\hat{S})\\[0.5em]
&~&+C_{vN}T^1(\hat{I})\cdot T^1(\hat{N}),
\end{array}\label{eq1}
\end{eqnarray}
in which the rotational constant $B_v=Y_{01}+Y_{11}(v+1/2)+Y_{21}(v+1/2)^2$, centrifugal distortion constant $D_v=-Y_{02}-Y_{12}(v+1/2)$, spin-rotational constant $\gamma_{vN}=\gamma_{00}+\gamma_{10}(v+1/2)+\gamma_{01}N(N+1)$, Fermi contact constant $b_F=b_v+c_v/3$ with hyperfine constant $b_v$ and dipole-dipole constant $c_v$, and the nuclear spin-rotational constant $C_{vN}$ is generally small enough, at the magnitude of kilohertz, to be neglected in our calculations, but listed here for completeness. The required Dunham coefficients and $\gamma, b_v, c_v$ parameters are listed in Table \ref{table3}.

\begin{table}[]
\caption{\label{table3}
The Dunham coefficients $Y$ (from Ref.\cite{Guo1995}), the spin-rotational constants $\gamma$ (from Ref.\cite{Ryzlewicz1980}) and hyperfine constants $b,c$ (from Ref.\cite{Ernst1986}) used in hyperfine structure calculation for $^{138}$BaF molecule. All values here in unit of MHz.}
\begin{ruledtabular}
\begin{tabular}{cccc}
Parameters & Values & Parameters & Values \\
\hline
$Y_{01}$ & 6491.3946 & $\gamma_{00}$ & 80.9840\\
$Y_{02}$ & $-5.5248\times10^{-3}$ & $\gamma_{10}$ & $-58.4\times10^{-3}$\\
$Y_{11}$ & $-34.8784$ & $\gamma_{01}$ & $0.112\times10^{-3}$\\
$Y_{12}$ & $-9.7632\times10^{-6}$  & $b_0$ & 63.509\\
$Y_{21}$ & $13.0288\times10^{-3}$ & $c_0$ & 8.224\\
\end{tabular}
\end{ruledtabular}
\end{table}

Since the good quantum numbers for Hund's case (b) state $X^2\Sigma_{1/2}$ are $N,J,F$, we expand the Hamiltonian (\ref{eq1}) under basis $\vert\phi\rangle=\vert N,S,J,I,F,m_F\rangle$. The corresponding matrix elements for each term of (\ref{eq1}) are,
\begin{eqnarray}
~&~&\langle\phi'\vert B_v\hat{N}^2-D_{v}\hat{N}^4 \vert\phi\rangle \nonumber\\[0.5em]
~&~&~=\delta_{N'N}\delta_{J'J}\delta_{F'F}\delta_{m'_Fm_F}\left(B_vN(N+1)-D_v[N(N+1)]^2\right),\label{eq2}
\\[1em]
~&~&\langle\phi'\vert\gamma_{v N}T^1(\hat{S})\cdot T^1(\hat{N})\vert\phi\rangle=\delta_{N'N}\delta_{J'J}\delta_{F'F}\delta_{m'_Fm_F}\gamma_{v N} \nonumber\\[0.5em]
~&~&~~\times(-1)^{N+J+S}\{S\}^{1/2}\{N\}^{1/2}\left\{\begin{array}{ccc}
S & N & J \\
N & S & 1
\end{array}\right\},\label{eq3}
\\[1em]
~&~&\langle\phi'\vert b_FT^1(\hat{I})\cdot T^1(\hat{S}) \vert\phi\rangle=\delta_{N'N}\delta_{F'F}\delta_{m'_Fm_F}b_F \nonumber\\[0.5em]
~&~&~~\times(-1)^{J'+F+I+J+N+1+S}[J']^{1/2}[J]^{1/2}\{S\}^{1/2}\{I\}^{1/2} \nonumber\\[0.5em]
~&~&~~\times\left\{\begin{array}{ccc}
I & J' & F \\
J & I & 1
\end{array}\right\}\left\{\begin{array}{ccc}
J & S & N \\
S & J' & 1
\end{array}\right\}, \label{eq4}
\end{eqnarray}
\begin{eqnarray}
~&~&\langle\phi'\vert c_vT^1_{q=0}(\hat{I})T^1_{q=0}(\hat{S}) \vert\phi\rangle=\delta_{N'N}\delta_{F'F}\delta_{m'_Fm_F}(-\sqrt{30}c_v/3) \nonumber\\[0.5em]
~&~&~~\times(-1)^{J'+F+I+N}[J']^{1/2}[J]^{1/2}\{S\}^{1/2}\{I\}^{1/2}(2N+1) \nonumber\\[0.5em]
~&~&~~\times\left(\begin{array}{ccc}
N & 2 & N \\
0 & 0 & 0
\end{array}\right)\left\{\begin{array}{ccc}
I & J' & F \\
J & I & 1
\end{array}\right\}\left\{\begin{array}{ccc}
J & J' & 1 \\
N & N & 2 \\
S & S & 1
\end{array}\right\}, \label{eq5}
\\[1em]
~&~&\langle\phi'\vert C_{vN}T^1(\hat{I})\cdot T^1(\hat{N}) \vert\phi\rangle=\delta_{N'N}\delta_{F'F}\delta_{m'_Fm_F}C_{vN} \nonumber\\[0.5em]
~&~&~~\times(-1)^{2J+F'+I+N'+1+S}[J']^{1/2}[J]^{1/2}\{N\}^{1/2}\{I\}^{1/2} \nonumber\\[0.5em]
~&~&~~\times\left\{\begin{array}{ccc}
I & J & F' \\
J' & I & 1
\end{array}\right\}\left\{\begin{array}{ccc}
N & J & S \\
J' & N' & 1
\end{array}\right\}, \label{eq6}
\end{eqnarray}
where $[x]^{1/2}=\sqrt{2x+1}$ and $\{x\}^{1/2}=\sqrt{x(x+1)(2x+1)}$.

\begin{figure}[]
\includegraphics[width= 0.4\textwidth]{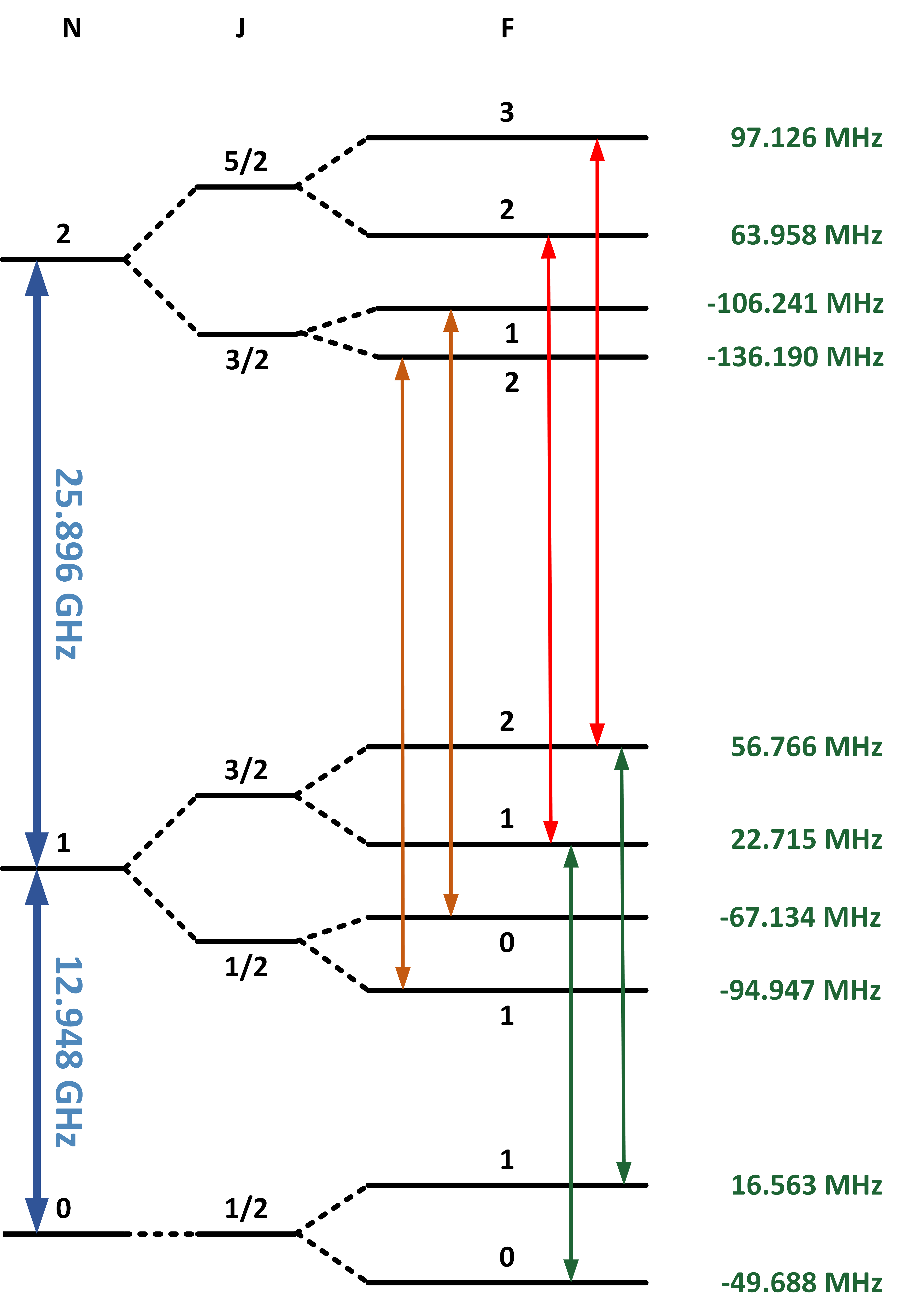}
\caption{\label{figure3}(Color online) The spin-rotational splittings and hyperfine levels for $N=0,1,2$ rotational states in $X^2\Sigma(v=0)$. The energy values are shown corresponding to the reference energy of each rotational state. The red, orange and green lines indicate the $\Delta J=+1, \Delta F=+1$ transitions, and the calculated values are listed in Table \ref{table4}, respectively.}
\end{figure}

\begin{figure}[t]
\includegraphics[width= 0.48\textwidth]{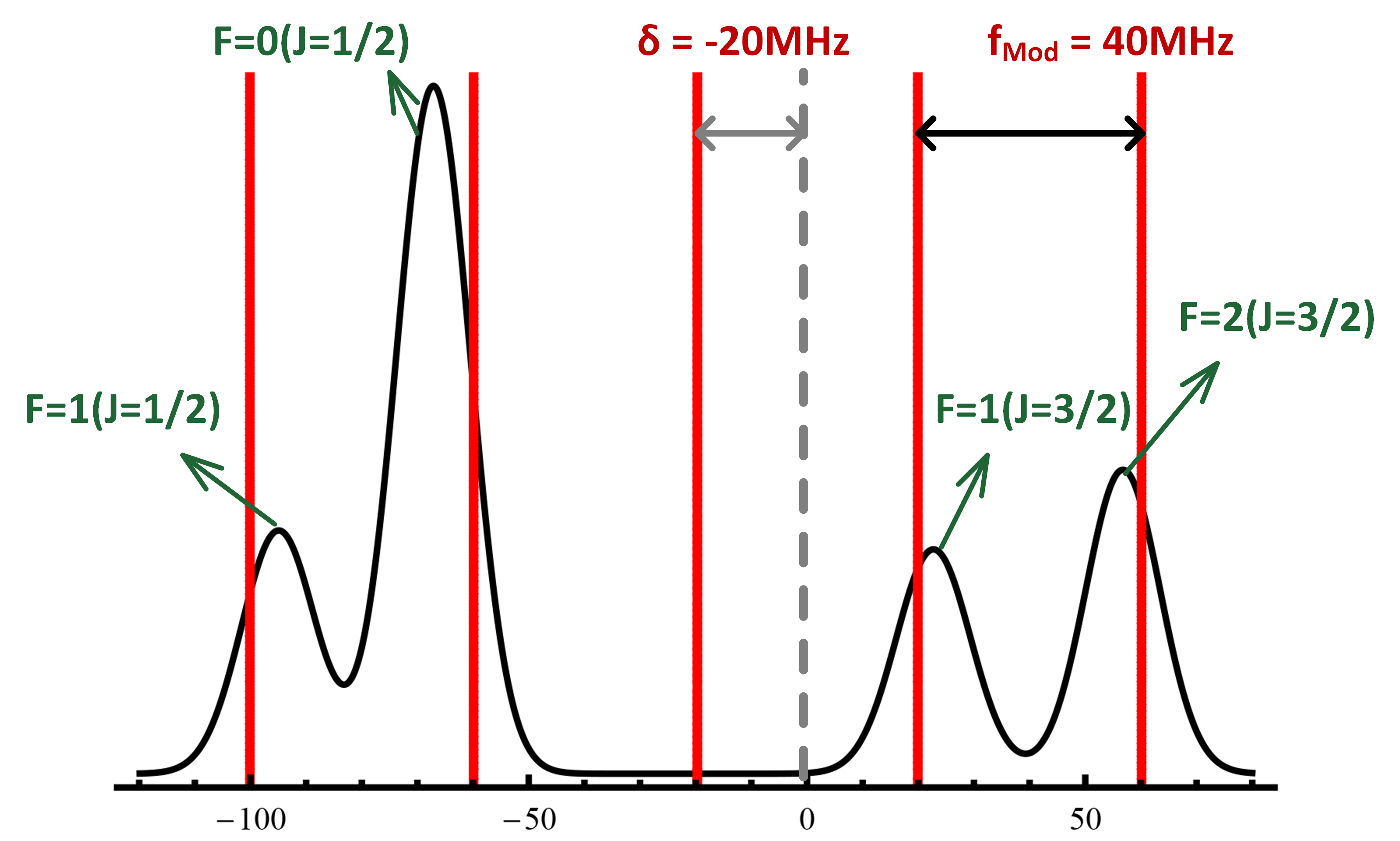}
\caption{\label{figure4}(Color online) The proposed sideband modulation scheme to the (re)pumping lasers to simutaneously cover all four hyperfine levels of $\vert X,N=1\rangle$. The calculated spectra (black line) are plotted with the branching ratio from the $\vert A, J=1/2, +\rangle$ to each hyperfine levels(see Fig.\ref{figure2}) as the strength of each peak, the calculated energy value in Fig.\ref{figure3} as the center frequency, and $\Gamma$ as the linewidth. The vertical red line indicate the sidebands of an EOM. By simply making the laser detuning $\delta=-20\text{MHz}$ and the modulated frequency $f_\text{Mod}=40$ MHz, the resultant sideband frequency values are matched within $\sim 3\Gamma$ detuned to the respective peaks.}
\end{figure}

By diagonalizing the $H_\text{eff}$ matrix, the energy splittings between different rotational hyperfine levels are obtained and illustrated in Fig.\ref{figure3}. The data therein is for $v=0$ case, while for higher $v=1,2$ states, the energy splittings differ rather small ($\sim$ kHz level) with those of $v=0$ because the $\gamma_{10}$ is at the magnitude of several tens of kHz. Based on the calculated data, we firstly discuss the sideband modulation to the pumping ($\lambda_{00}$) and repumping ($\lambda_{10}, \lambda_{21}$) lasers for $\vert X,N=1,F\rangle \leftrightarrow \vert A,J'=1/2, F'=0,1\rangle$ transitions. Figure \ref{figure4} shows the theoretically calculated fluorescence spectra and the proposed sideband frequency distributions generated by an electro-optical modulator (EOM). By simply choosing the laser detuning $\delta=-20~\text{MHz}$ and the modulating frequency $f_\text{Mod}=40~\text{MHz}$, the four hyperfine levels of $N=1$ are all addressed with detunings within $3\Gamma$ respectively. Here the laser detuning $\delta$ is defined by the frequency difference to the $\vert X,N=1\rangle \leftrightarrow\vert A, J'=1/2\rangle$ transition, while in SrF experiments\cite{Shuman2010,Barry2014} the laser detuning is experimentally determined
to be zero when maximal laser-induced fluorescence signal is obtained, and this zero value in turn serves as a benchmark to define the laser detunings.

\begin{table}[]
\caption{\label{table4}
Comparison of the calculated and experimentally observed transition frequencies for $\Delta J=J'-J=+1, \Delta F=F'-F=+1$ hyperfine transitions in rotational states of $X^2\Sigma_{1/2}$ state of BaF. The frequency difference $\Delta f=f_\text{calc.} - f_\text{obs.}$ is defined. The experimental data is taken from Ref.\cite{Ernst1986}.}
\begin{ruledtabular}
\begin{tabular}{cccccc}
$N'-N$ & $J'-J$ & $F'-F$ & $f_\text{calc.}$(MHz) & $f_\text{obs.}$(MHz) & $\Delta f$(kHz)\\
\hline
$1-0$ & $3/2-1/2$ & $1-0$ & 13020.2976 & 13020.286 & +11.6\\
~& ~& $2-1$ & 12988.0986 & 12988.110 & -11.4 \\[1em]
$2-1$ & $3/2-1/2$ & $1-0$ & 25856.5501 & 25865.572\footnotemark[1] & \\
~& ~ & $2-1$ & 25854.4139 & 25854.434 & -20.0\\[0.2em]
~& $5/2-3/2$ & $2-1$ & 25936.9012 & 25936.873 & +28.2\\
~ & ~& $3-2$ & 25936.0181 & 25936.006 & +12.1\\
\end{tabular}
\end{ruledtabular}
\footnotetext[1]{This value in Ref.\cite{Ernst1986} has significant difference with our calculated value.}
\end{table}

On the other hand, different with SrF molecule, an additional feature of BaF is the leakage decay to $\vert X, N=0,2, +\rangle$ states via the metastable $\Delta$ state (see Fig.\ref{figure2}), which leads to the unexpected rotational branchings. Fortunately, a microwave remixing method could perfectly solve this problem \cite{Yeo2015}. In Fig.\ref{figure3}, the  energy splittings for every neighboring two rotational states and the energy values for all hyperfine states are shown, and one can easily obtain the energy for every arbitrary possible transition. Here we plan to use the $\Delta J=+1,\Delta F=+1$ transitions to implement the microwave remixing, and the corresponding frequency values are listed in Table \ref{table4}. We have compared our calculated values with the experimentally observed spectra data in Ref.\cite{Ernst1986} and found that the differences are within several kHz. Such small difference also demonstrates the reliability of our calculations from another side. Microwave radiation tuned to $f_0=12948\pm15~\text{MHz}$ can drive $\vert N=0,F=0\rangle\leftrightarrow\vert N=1,J=1/2,F=0\rangle$ and $\vert N=0,F=1\rangle\leftrightarrow\vert N=1,J=3/2,F=1\rangle$ transitions to mix the $N=0$ hyperfine states with those in $N=1$, while the $N=2$ is remixed to $N=1$ just by doubling the frequency $f_0$ to drive $\Delta J=+1,\Delta F=+1$ transitions. The detunings for the six transitions are less than $10~\text{MHz}$.

Till now, we have discussed the quasi-closed optical cycling for BaF molecule in detail over all directions: the short lifetime of $A^2\Pi_{1/2}$ state, highly diagonal FCFs, parity selection rules and microwave remixing assisted rotational branching elimination, sideband modulation to adress all four hyperfine levels of $\vert X,N=1\rangle$. In following sections, we mainly focus on the molecular properties correlated with magneto-optical trapping experiment, including the branching ratios, energy splitting under external field and the lifetime estimation of the delta state.

\section{\label{section4}Branching ratios for \emph{A-X} transition}

The branching ratios reflect the distributions of the transition strengths for various all possible hyperfine decay paths. It is therefore necessary to calculate the branching ratios to determine the required laser intensities for certain transitions and for reproducing the molecular population distribution with experimentally observed line strengths to these hyperfine levels. In this section, we summarize the calculation details of the hyperfine branching ratios in $A-X$ transition for BaF following the derivations in Ref.\cite{Wall2008}.

Before deriving the matrix elements for electric dipole transition in $A-X$, we firstly investigate the $J$-mixing of the hyperfine levels in the ground $X$ state, where spin-rotation interaction produces $J$-splittings and then hyperfine interaction results in different $F$-branchings. The mixing coefficients are obtained by diagonalizing the $H_\text{eff}$ matrix; see Eq.(\ref{eq1}). For $N=0$, $J=1/2$, no mixing exists; while for $N=1$ or $N=2$ manifold, the hyperfine levels with same $F$ value but in different $J$ components suffer from the so-called $J$-mixing. Here we still label the nominal mixed hyperfine level as $\vert N,J,F\rangle$, but the pure $J$ state as $\vert N,J(F)\rangle$. The mixing situations for $N=1$ and $N=2$ are shown in Table \ref{table5}. Knowledge of $J$-mixing is required in following branching ratios calculations.

\begin{table}[]
\caption{\label{table5}
The $J$-mixing phenomena in $N=1$ and $N=2$ rotational states of ground $X$ state. The corresponding coefficients are
$\alpha_1=0.9593,\beta_1=0.2824$ and $\alpha_2=0.9858,\beta_2=0.1679$.}
\begin{ruledtabular}
\begin{tabular}{llr}
$N$ & mixed label & superposition of pure $J$ states \\
\hline
1 & $\vert J=1/2,F=0\rangle$ & $\vert J=1/2(F=0)\rangle$ \\
~ & $\vert J=1/2,F=1\rangle$ & $\alpha_1\vert J=1/2(F=1)\rangle + \beta_1\vert J=3/2(F=1)\rangle$ \\
~ & $\vert J=3/2,F=1\rangle$ & $-\beta_1\vert J=1/2(F=1)\rangle + \alpha_1\vert J=3/2(F=1)\rangle$ \\
~ & $\vert J=3/2,F=2\rangle$ & $\vert J=3/2(F=2)\rangle$ \\[0.5em]
2 & $\vert J=3/2,F=1\rangle$ & $\vert J=3/2(F=1)\rangle$ \\
~ & $\vert J=3/2,F=2\rangle$ & $\alpha_2\vert J=3/2(F=2)\rangle + \beta_2\vert J=5/2(F=2)\rangle$ \\
~ & $\vert J=5/2,F=2\rangle$ & $-\beta_2\vert J=3/2(F=2)\rangle + \alpha_2\vert J=5/2(F=2)\rangle$ \\
~ & $\vert J=5/2,F=3\rangle$ & $\vert J=5/2(F=3)\rangle$ \\
\end{tabular}
\end{ruledtabular}
\end{table}

Now let us discuss the calculation of the branching ratios for all possible hyperfine decays from $\vert A, J=1/2, +\rangle$ to $\vert X, N=1, -\rangle$. We firstly convert the nominal basis sets $\vert X; N,J,F\rangle$ and $\vert A; J, +\rangle$ into Hund's case (a) basis $\vert \Lambda,S,\Sigma,\Omega,J,I,F,m_F\rangle$. The nominal $J$-mixed $\vert N,J,F\rangle$ states in $X$ can be written as superpositions of pure $J$ states $\vert \Lambda; N,S,J(F)\rangle$ (abbreviated as $\vert N,J(F)\rangle$ in Table \ref{table5}). The pure $J$ state is Hund's case (b) state and can further be converted to case (a) basis as
\begin{eqnarray}
\vert \Lambda; N,S,J(F)\rangle&=&\sum\limits_{\Omega}\sum_{\Sigma}(-1)^{J+\Omega}\sqrt{2N+1}\nonumber\\[0.5em]
~&\times&\left(\begin{array}{ccc}
S & N & J \\
\Sigma & \Lambda & -\Omega
\end{array}\right)\vert\Lambda,S,\Sigma,\Omega,J,F\rangle,\label{eq7}
\end{eqnarray}
while for $A$ state,
\begin{eqnarray}
\vert |\Lambda|, J, \pm\rangle &=& \frac{1}{\sqrt{2}}\left(\vert \Lambda; S,\Sigma; J, \Omega\rangle\right. \nonumber\\[0.5em]
~&\pm&(-1)^{J-S}\left.\vert -\Lambda; S, -\Sigma; J, -\Omega\rangle\right).\label{eq8}
\end{eqnarray}

Then we calculate the matrix element for electric dipole transition between two Zeeman sublevels labeling as $\vert \psi_e\rangle$ and $\vert \psi_g\rangle$ under Hund's case (a) basis, that is,
\begin{eqnarray}
\langle d\rangle&=&\langle \psi_e\vert T^1_p(\hat{d})\vert\psi_g\rangle \nonumber\\[0.5em]
~&=&\langle\alpha_e; J_e,I_e,F_e,m_{F,e}\vert T^1_p(\hat{d})\vert \alpha_g;J_g,I_g,F_g,m_{F,g}\rangle \nonumber\\[0.5em]
~&=&(-1)^{F_e-m_{F,e}+F_g+J_e+I_g+1}[F_e]^{1/2}[F_g]^{1/2} \nonumber\\[0.5em]
~&~&\times\left(\begin{array}{ccc} F_e & 1 & F_g \\ -m_{F,e} & p & m_{F,g}\end{array}\right)\left\{\begin{array}{ccc} J_g & F_g & I_g \\ F_e & J_e & 1\end{array}\right\} \nonumber\\[0.5em]
~&~&\times\langle\alpha_e; J_e\vert\vert T^1(\hat{d})\vert\vert\alpha_g;J_g\rangle, \label{eq9}
\end{eqnarray}
where $\vert \alpha\rangle=\vert \Lambda; S,\Sigma; \Omega\rangle$. Applying the Wigner-Eckart theorem to the last term in Eq.(\ref{eq9}), we obtain
\begin{eqnarray}
~&~&\langle\Lambda_e;S_e,\Sigma_e;\Omega_e,J_e ||T^1(\hat{d})||\Lambda_g;S_g,\Sigma_g;\Omega_g,J_g\rangle \nonumber\\[0.5em]
~&~&=\sum\limits_{q=-1}^{1}(-1)^{J_e-\Omega_e}[J_e]^{1/2}[J_g]^{1/2}\left(\begin{array}{ccc} J_e & 1 & J_g \\ -\Omega_e & q & \Omega_g \end{array}\right) \nonumber \\[0.5em]
~&~&~~~~\times\langle\Lambda_e;S_e,\Sigma_e ||T_q^1(\hat{d}) ||\Lambda_g;S_g,\Sigma_g\rangle, \label{eq10}
\end{eqnarray}
where $\Sigma_e=\Sigma_g$ should be satisfied since the electrical dipole operator $T^1_q(\hat{d})$ can neither change the electron spin nor the spin projection, and the matrix element $\langle \Lambda_e||T_q^1||\Lambda_g\rangle$ is common for all $\Delta\Lambda=\pm 1$ transitions.

Putting Eqs.(\ref{eq7})-(\ref{eq10}) and the formula in Table \ref{table5} all together, the branching ratios for decays from hyperfine Zeeman sublevels in $\vert A, J=1/2, +\rangle$ to those in $\vert X, N=1,-\rangle$ state are obtained and summarized in Table \ref{table6}. The calculated branching ratios provide instructive directions to future laser cooling and trapping experiments on BaF molecule.

\begin{table}[b]
\caption{\label{table6}
Calculated hyperfine branching ratios for decays from $\vert A,J'=1/2,+\rangle$ state to $\vert X, N=1,-\rangle$ state. The line strengths in Fig.\ref{figure4} are estimated with the values here.}
\begin{ruledtabular}
\begin{tabular}{ccr|cccc}
~ & ~ & ~ & $F'=0$ & ~ & $F'=1$ & ~ \\
$J$ & $F$ & $m_F$ & $m'_F=0$ & $m'_F=-1$ & $m'_F=0$ & $m'_F=1$ \\
\hline
1/2 & 0 & 0 & 0 & 2/9 & 2/9 & 2/9 \\[0.5em]
~ & ~ & $-1$ & 0.2985 & 0.1641 & 0.1641 & 0 \\
1/2 & 1 & 0 & 0.2985 & 0.1641 & 0 & 0.1641 \\
~ & ~ & 1 & 0.2985 & 0 & 0.1641 & 0.1641 \\[0.5em]
~ & ~ & -1 & 0.0348 & 0.0859 & 0.0859 & 0 \\
3/2 & 1 & 0 & 0.0348 & 0.0859 & 0 & 0.0859 \\
~ & ~ & 1 & 0.0348 & 0 & 0.0859 & 0.0859 \\[0.5em]
~ & ~ & -2 & 0 & 1/6 & 0 & 0 \\
~ & ~ & -1 & 0 & 1/12 & 1/12 & 0 \\
3/2 & 2 & 0 & 0 & 1/36 & 1/9 & 1/36 \\
~ & ~ & 1 & 0 & 0 & 1/12 & 1/12 \\
~ & ~ & 2 & 0 & 0 & 0 & 1/6 \\
\end{tabular}
\end{ruledtabular}
\end{table}

\begin{table}[]
\caption{\label{table7}
Calculated hyperfine branching ratios for decays from $\vert A,J'=1/2,-\rangle$ state to $\vert X, N=0,2,+\rangle$ state.}
\begin{ruledtabular}
\begin{tabular}{cccr|cccc}
~ & ~ & ~ & ~ & $F'=0$ & ~ & $F'=1$ & ~ \\
$N$ & $J$ & $F$ & $m_F$ & $m'_F=0$ & $m'_F=-1$ & $m'_F=0$ & $m'_F=1$ \\
\hline
0 & 1/2 & 0 & 0 & 0 & 2/9 & 2/9 & 2/9 \\[1em]
~ & ~ & ~ & $-1$ & 2/9 & 2/9 & 2/9 & 0 \\
~ & ~ & 1 & 0 & 2/9 & 2/9 & 0 & 2/9 \\
~ & ~ & ~ & 1 & 2/9 & 0 & 2/9 & 2/9 \\
\hline
~ & ~ & ~ & -1 & 1/9 & 1/36 & 1/36 & 0 \\
2 & 3/2 & 1 & 0 & 1/9 & 1/36 & 0 & 1/36 \\
~ &~ & ~ & 1 & 1/9 & 0 & 1/36 & 1/36 \\[1em]
~&~ & ~ & -2 & 0 & 0.1620 & 0 & 0 \\
~&~ & ~ & -1 & 0 & 0.0810 & 0.0810 & 0 \\
~&3/2 & 2 & 0 & 0 & 0.0270 & 0.1080 & 0.0270 \\
~&~ & ~ & 1 & 0 & 0 & 0.0810 & 0.0810 \\
~&~ & ~ & 2 & 0 & 0 & 0 & 0.1620 \\[1em]
~ & ~ & ~ & -2 & 0 & 0.0047 & 0 & 0 \\
~ & ~ & ~ & -1 & 0 & 0.0023 & 0.0023 & 0 \\
~ & 5/2 & 2 & 0 & 0 & 0.0008 & 0.0031 & 0.0008 \\
~ & ~ & ~ & 1 & 0 & 0 & 0.0023 & 0.0023 \\
~ & ~ & ~ & 2 & 0 & 0 & 0 & 0.0047 \\
\end{tabular}
\end{ruledtabular}
\end{table}

\section{\label{section5}Zeeman splittings under magnetic field}

\begin{figure}[]
\includegraphics[width= 0.5\textwidth]{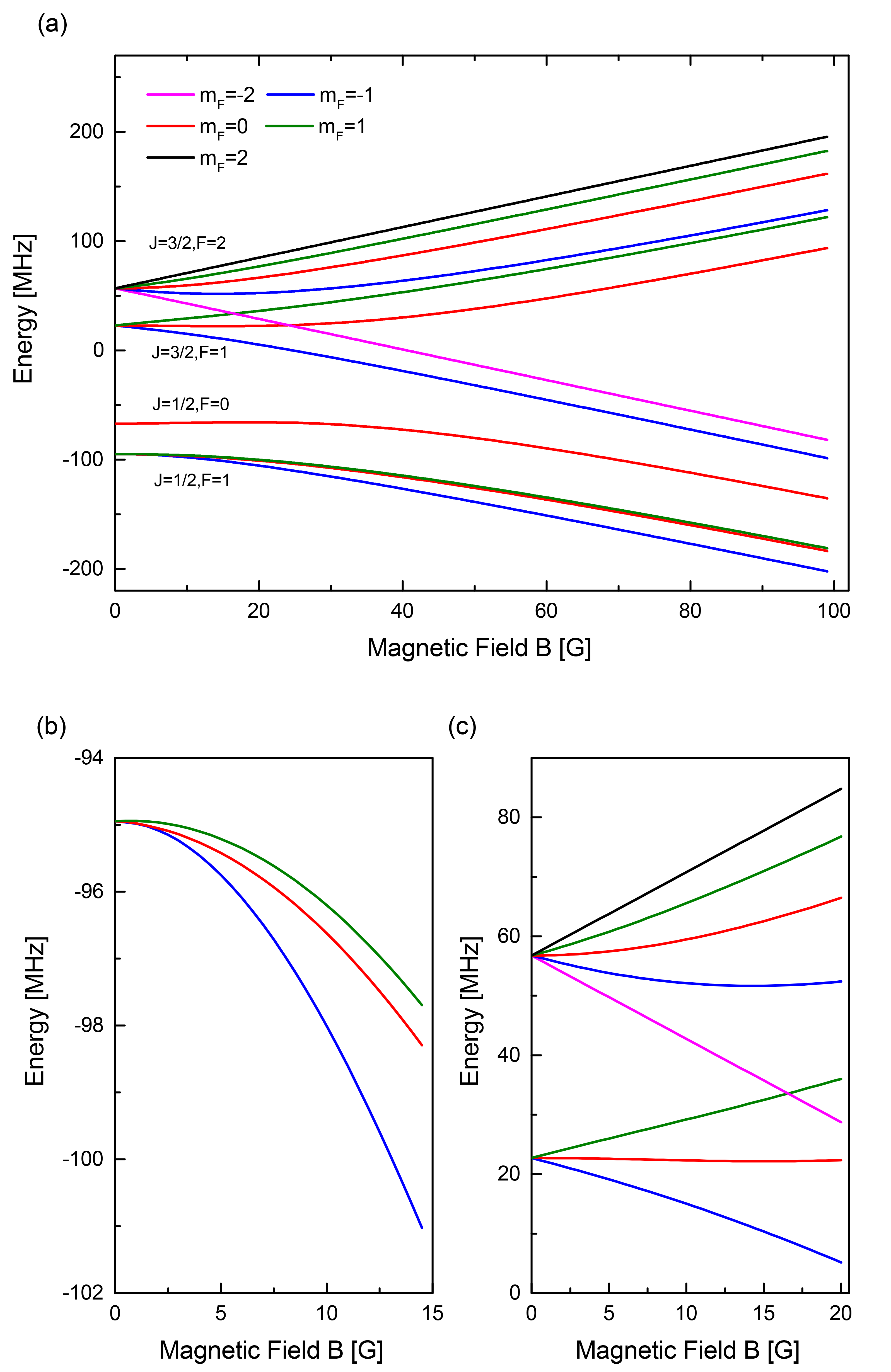}
\caption{\label{figure5}(Color online) Zeeman splittings for sublevels in the four hyperfine states of $\vert X,N=1,-\rangle$ manifold. (a) The whole picture, sublevels labeled by same $m_F$ quantum number are plotted in same color. (b) and (c) are zoom-in plots of $\vert J=1/2,F=1\rangle$ and $\vert J=3/2,F=1,2\rangle$ manifolds respectively at small magnetic field region. }
\end{figure}

To achieve the remixing of dark Zeeman sublevels \cite{Barry2014} and further trapping of molecules \cite{Hummon2013,Norrgard2016}, an external magnetic field is usually applied to the molecules. In a magneto-optical trap, the magnetic field is employed to create a position dependent restoring force to cool and confine the atoms or molecules. The degeneracy of the total $2F+1$ Zeeman sublevels within a hyperfine state of quantum number $F$ is destroyed by the external magnetic field. In this section, we focus on the Zeeman splittings of the two states involved with the cooling transition $\vert X, N=1, -\rangle \leftrightarrow \vert A, J=1/2,+\rangle$. Under the external field $B_z$, the Zeeman Hamiltonian
\begin{equation}
H_z=\left[g_S\mu_BT^1_{p=0}(\hat{S})+g_L\mu_BT^1_{p=0}(\hat{L})-g_I\mu_NT^1_{p=0}(\hat{I})\right]B_z, \label{eq11}
\end{equation}
together with the spin-rotation and hyperfine interaction terms in Eq.(\ref{eq1}) contribute to the splittings of Zeeman sublevels. For the ground state $\vert X, N=1,-\rangle$, the three terms in Hamiltonian $H_z$ are expanded under basis $\vert \phi\rangle=\vert N,S,J,I,F,m_F\rangle$ respectively as
\begin{eqnarray}
~& ~& \langle \phi \vert g_S\mu_BT^1_{p=0}(\hat{S})\vert\phi'\rangle = g_S\mu_B(-1)^{F-m_F}\left(\begin{array}{ccc} F & 1 & F' \\ -m_F & 0 & m'_F \end{array}\right) \nonumber \\[0.5em]
~& ~& ~~~\times (-1)^{F'+J+1+I}[F']^{1/2}[F]^{1/2} (-1)^{J'+N+1+S}[J']^{1/2}[J]^{1/2} \nonumber \\[0.5em]
~& ~& ~~~\times \{S\}^{1/2} \left\{\begin{array}{ccc} F & J & I \\ J' & F' & 1\end{array}\right\}\left\{\begin{array}{ccc} J & S & N \\ S & J' & 1\end{array}\right\}, \label{eq12}
\end{eqnarray}

\begin{eqnarray}
~& ~& \langle \phi \vert g_L\mu_BT^1_{p=0}(\hat{L})\vert\phi'\rangle = g_L\mu_B(-1)^{F-m_F}\left(\begin{array}{ccc} F & 1 & F' \\ -m_F & 0 & m'_F \end{array}\right) \nonumber \\[0.5em]
~& ~& ~~~\times (-1)^{F'+J+1+I}[F']^{1/2}[F]^{1/2} (-1)^{J'+N+1+S}[J']^{1/2}[J]^{1/2} \nonumber \\[0.5em]
~& ~& ~~~\times (-1)^{N-\Lambda} [N]^{1/2}[N']^{1/2} \left\{\begin{array}{ccc} F & J & I \\ J' & F' & 1\end{array}\right\}\left\{\begin{array}{ccc} J & N & S \\ N' & J' & 1\end{array}\right\} \nonumber \\[0.5em]
~& ~& ~~~\times \left(\begin{array}{ccc} N & 1 & N' \\ -\Lambda& 0 & \Lambda \end{array}\right)\Lambda, \label{eq13}
\end{eqnarray}
and
\begin{eqnarray}
~& ~& \langle \phi \vert g_I\mu_BT^1_{p=0}(\hat{I})\vert\phi'\rangle = g_I\mu_N\delta_{J,J'}(-1)^{F-m_F}\left(\begin{array}{ccc} F & 1 & F' \\ -m_F & 0 & m'_F \end{array}\right) \nonumber \\[0.5em]
~& ~& ~~~\times (-1)^{F'+J+1+I}[F']^{1/2}[F]^{1/2} \{I\}^{1/2} \left\{\begin{array}{ccc} F & I & J \\ I & F' & 1\end{array}\right\}. \label{eq14}
\end{eqnarray}
Since $\Lambda=0$ for $X^2\Sigma$ state, the second term (\ref{eq13}) vanishes; and the third term (\ref{eq14}) is smaller enough to be neglected due to $\mu_B/\mu_N\approx 1836$. Consequently, the effective Hamiltonian matrix is constructed by the elements from Eqs.(\ref{eq3})-(\ref{eq5}) and (\ref{eq12}). By diagonalizing the matrix at different magnetic field strengths, we obtain the Zeeman energy splittings as shown in Fig.\ref{figure5}. The splitting behaviors at weak and strong magnetic field strengths show different features. In strong field region, the sublevels with different $F, m_F$ values totally split.

However, typical magneto-optical trap for molecules employs field of about only several Gauss strength. When magnetic field is weak, the Zeeman Hamiltonian is just a perturbation to the hyperfine interaction term, thus the energy shifts for sublevles in $\vert J=3/2,F=1,2\rangle$ and $\vert J=1/2,F=0\rangle$ show nearly linear variations, as illustrated in Fig.\ref{figure5}(c). But specially, for $\vert J=1/2, F=1\rangle$ manifold, the shifts do not linearly variate along with the field strengths, that is, the linear region is rather small (less than 5 Gauss); see Fig.\ref{figure5}(b). The reason is that the matrix element of the hyperfine term for $\vert J=1/2,F=1\rangle$ is relatively smaller than those for $\vert J=3/2,F=1,2\rangle$, and thus much more easily be perturbed by the external magnetic field. Another point is that the variation behavior can not be correctly described by the typical $g$-factors, instead mixed $g$-factors are used to describe the linear gradients of the Zeeman shifts. By applying a rather small magnetic field strength, say 0.01 Gauss, to the Zeeman Hamiltonian (\ref{eq11}), we obtain the energy differences with the degenerate hyperfine levels at zero strength for all Zeeman sublevels, and the mixed $g$-factors are yielded by a transformation.

\begin{table}[]
\caption{\label{table8}
The typical $g$-factors ($g_F$) and the mixed ones for the four hyperfine levels in $\vert X,N=1,-\rangle$ state.}
\begin{ruledtabular}
\begin{tabular}{lcr}
 state label & typical $g$-factor $g_F$ & mixed $g$-factor \\
\hline
 $\vert J=1/2,F=0\rangle$ & 0.00 & 0.000 \\
 $\vert J=1/2,F=1\rangle$ & -0.33 & 0.015 \\
 $\vert J=3/2,F=1\rangle$ & 0.83 & 0.485 \\
 $\vert J=3/2,F=2\rangle$ & 0.50 & 0.500
\end{tabular}
\end{ruledtabular}
\end{table}

Table \ref{table8} gives the typical $g$-factors and the mixed $g$-factors for the four hyperfine states in $\vert X,N=1,-\rangle$. For states $\vert J=1/2,F=0\rangle$ and $\vert J=3/2,F=2\rangle$, no $J$-mixing exists, and both the $g$-factors keep identical with those typical $g_F$ for pure $J$ states. However, the mixed $g$-factors for two $\vert F=1\rangle$ states change significantly due to the $J$-mixing effect which is much stronger than the Zeeman effect from external field. Different with SrF \cite{Barry2014} and MgF \cite{Xu2016} where mixed $g$-factor for $\vert J=1/2,F=1\rangle$ is negative and an additional opposite polarized trapping laser should be combined, the $g$-factor for BaF here is positive, that is, all hyperfine manifolds have $g>0$ and thus laser addressing different manifolds could take same polarizations for optimal trapping.

However, Tarbutt \cite{Tarbutt2015} has pointed out that even for Type-II magneto-optical trap the restoring force and the laser polarizations are still mainly determined by the $g$-factor of the excited state. Under magnetic field strength $B_z$, the Zeeman interaction energy for Hund's case (a) state $\vert \Lambda; J, \Omega; F, m_F\rangle$ is given by
\begin{equation}
\Delta E_z=\frac{g_L\Lambda+g_s\Sigma}{J(J+1)}\Omega\mu_BB_zm_F, \label{eq15}
\end{equation}
where $g_L\approx 1$, $g_S=2.0023$. For $\vert A^2\Pi,J=1/2\rangle$ state, $\Delta E_z=7.7\times 10^{-4}\mu_Bm_FB_z$. The $g$-factor is close to zero.

Fortunately, the excited $\Pi_{1/2}$ state in fact is not totally a pure state, and it is usually spin-orbital mixed with upper $\Sigma_{1/2}$ state, which introduces additional parity-dependent terms to the Zeeman Hamiltonian \cite{Brown2003}. According to the derivations in Ref.\cite{Brown2003}, for $\vert A, J=1/2, \pm, F=1\rangle$ state, the parity-dependent Zeeman shift is
\begin{equation}
\Delta E_{pz\pm}=\pm g_p\mu_Bm_FB_z, \label{eq16}
\end{equation}
where $-/+$ indicate the odd/even parity, and $g_p$ is derived from $\Lambda$-doubling coefficients and the rotational constant \cite{Tarbutt2015}. For BaF molecule in $\vert A,J=1/2,+,F=1\rangle$, $g_p=\frac{1}{3}(p+2q)/(2B_e)=-0.202$, which is much larger than $g_p=-0.088$ for SrF \cite{Steimle1978}, $g_p=-0.065$ for YO \cite{Hummon2013} and $g_p=-0.021$ for CaF \cite{Nakagawa1978}. Larger $g$-factor of the upper state should result in a larger magneto-optical trapping force, and one can also resort to the rapid switching of the polarization of the trapping laser to achieve better trapping effect.

\section{\label{section6} The Delta state}

In Sec.\ref{section3}, we have discussed the undesired leakage decay from $\vert A,J=1/2,+\rangle$ to $\vert X,N=0,2,+\rangle$ via the $A'^2\Delta_{3/2}$ state, which will break the quasi-closed optical cycling. One concerned problem for our future laser cooling experiment is the decay rate from the $\vert A', J=3/2,-\rangle$ to the $\vert X,N=0,2,+\rangle$ state, that is, the lifetime of the metastable $\Delta$ state. In fact, the $\Delta$ state decay back to $X$ state due to its mixing with upper $A^2\Pi$ state and the selection rules forbid direct electric dipole transition from $\Delta$ to $\Sigma$ where $\Delta\Lambda=2$. In this section, we focus on the identification of the mixing between the $A'^2\Delta_{3/2}$ and $A^2\Pi$ states, and then make an estimation of the lifetime of the $A'^2\Delta$ state based on the mixing coefficients just like that performed in YO molecule \cite{Collopy2015}.

The mixing of $\Delta_{3/2}$ and $\Pi_{1/2,3/2}$ states originates from the spin-orbit and rotational electronic Coriolis interactions, denoted as $H_\text{so}$ and $H_\text{cor}$ respectively.
The spin-orbit operator $H_\text{so}=\sum_i\sum_{q=\pm 1}T^1_q(\hat{a}_i\hat{l}_i)T^1_{-q}(\hat{s}_i)$ describes the total effect from the interaction of each electron's spin ($\hat{s}_i$) with its own motion (described by the angular momentum $\hat{l}_i$), while the Coriolis operator $H_\text{cor}=-\frac{\hbar}{4\pi c\mu r^2}\sum_{q=\pm 1}T^1_q(\hat{J})T_{-q}^1(\hat{L})$ reflects the effect from the interaction between electron's motion and molecular rotation. Both the operators can couple two states with $\Delta\Lambda=\pm 1$.
The detailed descriptions for the two operators are discussed in Refs.\cite{Chalek1976,Brown2003}. Following the procedure in Ref.\cite{Chalek1976}, the off-diagonal nonzero matrix elements for $H_\text{so}$ and $H_\text{cor}$ operators are given as
\begin{eqnarray}
\begin{array}{lcl}
\langle v, A'^2\Delta_{3/2}\vert H_\text{cor}\vert A^2\Pi_{1/2},v'\rangle &=& -\sqrt{3}B_{vv'}b_2, \\[0.5em]
\langle v, A'^2\Delta_{3/2}\vert H_\text{cor}\vert A^2\Pi_{3/2},v'\rangle &=& B_{vv'}b_2,\\[0.5em]
\langle v, A'^2\Delta_{3/2}\vert H_\text{so}\vert A^2\Pi_{3/2},v'\rangle &=& \langle v\vert v'\rangle a_2,\\[0.5em]
\langle v, A'^2\Pi_{1/2}\vert H_\text{cor}\vert A^2\Pi_{3/2},v'\rangle &=& -\sqrt{3}B_{vv'},
\end{array} \label{17}
\end{eqnarray}
where $a_2=A_e^{\Pi}$ (see Table \ref{table1}), $b_2=2$ \cite{Chalek1976}, $\langle v\vert v'\rangle$ is the vibrational wavefunction overlap, and $B_{vv'}=\frac{\hbar}{4\pi c \mu}\langle v\vert (1/r^2)\vert v'\rangle$ is a vibrational averaging (over internuclear distance $r$) value. The relevant $\langle v\vert v'\rangle$ and $B_{vv'}$ values for $\vert A'^2\Delta_{3/2},v=0,1,2\rangle$, $\vert A^2\Pi_{1/2},v=0,1,2\rangle$ and $\vert A^2\Pi_{3/2},v=0,1,2\rangle$ states are listed in Table \ref{table9} and Table \ref{table10} respectively. The vibrational wavefunctions are evaluated with the method in Sec.\ref{section2}.

\begin{table}[]
\caption{\label{table9}
The vibrational wavefunction overlaps $\langle v\vert v'\rangle$ for $A'^2\Delta_{3/2}-A^2\Pi_{1/2}$ and $A'^2\Delta_{3/2}-A^2\Pi_{3/2}$.}
\begin{ruledtabular}
\begin{tabular}{ccccc}
 ~ & ~ & ~ & $A'^2\Delta_{3/2}(v)$ & ~ \\
 ~ & ~ & 0 & 1 & 2 \\
\hline
 ~ & 0 & 0.99228 & -0.12190 & 0.02235  \\
 $A^2\Pi_{3/2}(v')$ & 1 & 0.12401 & 0.97733 & -0.16722 \\
 ~ & 2 & -0.00135 & 0.17306 & 0.96321 \\[0.5em]
 ~ & 0 & 0.97511 & 0.21781 & 0.041001  \\
 $A^2\Pi_{1/2}(v')$ & 1 & -0.21962 & 0.92411 & 0.30419 \\
 ~ & 2 & 0.03027 & -0.30913 & 0.87070
\end{tabular}
\end{ruledtabular}
\end{table}

\begin{table}[]
\caption{\label{table10}
The $B_{vv'}$ values for $A'^2\Delta_{3/2}-A^2\Pi_{1/2}$, $A'^2\Delta_{3/2}-A^2\Pi_{3/2}$ and $A^2\Pi_{1/2}-A^2\Pi_{3/2}$.}
\begin{ruledtabular}
\begin{tabular}{ccccc}
 ~ & ~ & ~ & $A'^2\Delta_{3/2}(v)$ & ~ \\
 ~ & ~ & 0 & 1 & 2 \\
\hline
 ~ & 0 & 0.20904 & -0.03477 & 0.00710  \\
 $A^2\Pi_{3/2}(v')$ & 1 & 0.01689 & 0.20466 & -0.04771 \\
 ~ & 2 & -0.00102 & 0.02336 & 0.20048 \\[0.5em]
 ~ & 0 & 0.20904 & -0.03473 & 0.00730  \\
 $A^2\Pi_{1/2}(v')$ & 1 & 0.01689 & 0.20467 & -0.04761 \\
 ~ & 2 & -0.00125 & 0.02334 & 0.20049 \\
\hline\hline
 ~ & ~ & ~ & $A'^2\Pi_{3/2}(v)$ & ~ \\
 ~ & ~ & 0 & 1 & 2 \\
\hline
 ~ & 0 & 0.21183& -0.00931& 0.00109  \\
 $A^2\Pi_{1/2}(v')$ & 1 & -0.00933& 0.21057& -0.01306 \\
 ~ & 2 & 0.00066& -0.01313& 0.20931
\end{tabular}
\end{ruledtabular}
\end{table}

We construct the Hamiltonian matrix by taking the eigenenergy values of the vibrational states (adding the energy gaps $T_e(\Pi_{3/2})-T_e(\Delta_{3/2})$ and $T_e(\Pi_{1/2})-T_e(\Delta_{3/2})$ for corresponding states in $A^2\Pi_{3/2}$ and $A^2\Pi_{1/2}$) as the diagonal elements. However, the eigenenergy arrays should be eigenvalues of the mixing Hamiltonian matrix, thus we should vary the diagonal elements one by one to make the eigenvalues of the mixing matrix closer to the eigenenergy values of those states involved step by step. When the eigenvalues are approaching closely to each respective eigenenergy for the nine vibrational states, the eigenvector can approximately describe the mixing of the pure Born-Oppenheimer $\Pi$ and $\Delta$ states. For $v=0$ state in mixed $A'^2\Delta_{3/2}$, the wavefunction is given by
\begin{eqnarray}
\vert \widetilde{\Delta_{3/2}},v=0\rangle &\approx & 0.88732\vert \Delta_{3/2},v=0\rangle - 0.45945\vert \Pi_{3/2},v=0\rangle \nonumber \\[0.5em]
~ & -&0.03539\vert \Pi_{3/2},v=1\rangle + 0.00072\vert \Pi_{3/2},v=2\rangle \nonumber \\[0.5em]
~ & + & 0.00052\vert \Pi_{1/2},v=0\rangle+.... \label{eq18}
\end{eqnarray}

The mixing coefficients here along with the vibrational wavefunction overlaps between $A^2\Pi$ and $X^2\Sigma$ can give an approximate estimation on the vibrational branching ratios from $A'^2\Delta_{3/2}$ to $X^2\Sigma$ states. For $\vert A',v=0\rangle$ state, about 91.2\% decay back to $\vert X, v=0\rangle$, 8.3\% to $\vert X,v=1\rangle$ and the remaining 0.4\% to $\vert X,v\ge 2\rangle$ states. In actual cooling procedure, rapid decay rates from $\vert A'^2\Delta_{3/2},J'=3/2,-\rangle$ to $\vert X, N=0,2,+\rangle$ are required, so the lifetime of the $\Delta$ state should not be too long. Considering the 21\% mixing with $A^2\Pi_{3/2}$ and the lifetime of $A^2\Pi_{3/2}$ state is 46 ns \cite{Berg1998}, the $A'^2\Delta_{3/2}$ has a lifetime of $\sim$ 220 ns, which guarantees that after the pumped molecules in excited $\vert A^2\Pi_{1/2},J=1/2,+\rangle$ relaxing to the $\vert A'^2\Delta_{3/2},J=3/2,-\rangle$ state, they can rapidly decay back to $\vert X,N=0,2,+\rangle$ states, and then be remixed back to $\vert X,N=1,-\rangle$ in optical cycling by the microwave scheme described in Sec.\ref{section3}.

\section{\label{section7}Conclusion}

To conclude, we have investigated the molecular structures and branching ratios for BaF molecule, and further demonstrated the feasibility of laser cooling and trapping. BaF molecule has both the identical properties with laser-cooled SrF and YO molecule and its own unique character. The short lifetime for the $A^2\Pi_{1/2}$ state, the highly diagonal FCFs and the microwave remixing of different rotational manifolds in $X$ states implement the qusi-closed optical cycling procedure. Based on our calculations on the hyperfine splittings, we have proposed a sideband modulation scheme to simultaneously pump the four hyperfine levels of $\vert X,N=1\rangle$ and an optimal choice of the microwave frequencies. The results of the branching ratios, Zeeman splittings and $g$-factors will serve as a reference for adjustment of laser powers and polarizations. Finally, we have checked again that the leakage process via $\Delta$ state can be eliminated with microwave remixing since the lifetime of the $\Delta$ state is estimated as about 220 ns. The results and proposed schemes show the feasibility of future laser cooling and trapping experiments on BaF molecule.

\begin{acknowledgments}

The authors acknowledge the support from the National Natural Science Foundation of China under grants 91636104, the Fundamental Research Funds for the Central Universities 2016QNA3007.

\end{acknowledgments}

\bibliography{BaF_info_doc}

\begin{thebibliography}{46}%
\makeatletter
\providecommand \@ifxundefined [1]{%
 \@ifx{#1\undefined}
}%
\providecommand \@ifnum [1]{%
 \ifnum #1\expandafter \@firstoftwo
 \else \expandafter \@secondoftwo
 \fi
}%
\providecommand \@ifx [1]{%
 \ifx #1\expandafter \@firstoftwo
 \else \expandafter \@secondoftwo
 \fi
}%
\providecommand \natexlab [1]{#1}%
\providecommand \enquote  [1]{``#1''}%
\providecommand \bibnamefont  [1]{#1}%
\providecommand \bibfnamefont [1]{#1}%
\providecommand \citenamefont [1]{#1}%
\providecommand \href@noop [0]{\@secondoftwo}%
\providecommand \href [0]{\begingroup \@sanitize@url \@href}%
\providecommand \@href[1]{\@@startlink{#1}\@@href}%
\providecommand \@@href[1]{\endgroup#1\@@endlink}%
\providecommand \@sanitize@url [0]{\catcode `\\12\catcode `\$12\catcode
  `\&12\catcode `\#12\catcode `\^12\catcode `\_12\catcode `\%12\relax}%
\providecommand \@@startlink[1]{}%
\providecommand \@@endlink[0]{}%
\providecommand \url  [0]{\begingroup\@sanitize@url \@url }%
\providecommand \@url [1]{\endgroup\@href {#1}{\urlprefix }}%
\providecommand \urlprefix  [0]{URL }%
\providecommand \Eprint [0]{\href }%
\providecommand \doibase [0]{http://dx.doi.org/}%
\providecommand \selectlanguage [0]{\@gobble}%
\providecommand \bibinfo  [0]{\@secondoftwo}%
\providecommand \bibfield  [0]{\@secondoftwo}%
\providecommand \translation [1]{[#1]}%
\providecommand \BibitemOpen [0]{}%
\providecommand \bibitemStop [0]{}%
\providecommand \bibitemNoStop [0]{.\EOS\space}%
\providecommand \EOS [0]{\spacefactor3000\relax}%
\providecommand \BibitemShut  [1]{\csname bibitem#1\endcsname}%
\let\auto@bib@innerbib\@empty
\bibitem [{\citenamefont {Carr}\ \emph {et~al.}(2009)\citenamefont {Carr},
  \citenamefont {DeMille}, \citenamefont {Krems},\ and\ \citenamefont
  {Ye}}]{Carr2009}%
  \BibitemOpen
  \bibfield  {author} {\bibinfo {author} {\bibfnamefont {L.~D.}\ \bibnamefont
  {Carr}}, \bibinfo {author} {\bibfnamefont {D.}~\bibnamefont {DeMille}},
  \bibinfo {author} {\bibfnamefont {R.~V.}\ \bibnamefont {Krems}}, \ and\
  \bibinfo {author} {\bibfnamefont {J.}~\bibnamefont {Ye}},\ }\href
  {http://stacks.iop.org/1367-2630/11/i=5/a=055049} {\bibfield  {journal}
  {\bibinfo  {journal} {New J. Phys.}\ }\textbf {\bibinfo {volume} {11}},\
  \bibinfo {pages} {055049} (\bibinfo {year} {2009})}\BibitemShut {NoStop}%
\bibitem [{\citenamefont {Yan}\ \emph {et~al.}(2013)\citenamefont {Yan},
  \citenamefont {Moses}, \citenamefont {Gadway}, \citenamefont {Covey},
  \citenamefont {Hazzard}, \citenamefont {Rey}, \citenamefont {Jin},\ and\
  \citenamefont {Ye}}]{Yan2013}%
  \BibitemOpen
  \bibfield  {author} {\bibinfo {author} {\bibfnamefont {B.}~\bibnamefont
  {Yan}}, \bibinfo {author} {\bibfnamefont {S.~A.}\ \bibnamefont {Moses}},
  \bibinfo {author} {\bibfnamefont {B.}~\bibnamefont {Gadway}}, \bibinfo
  {author} {\bibfnamefont {J.~P.}\ \bibnamefont {Covey}}, \bibinfo {author}
  {\bibfnamefont {K.~R.~A.}\ \bibnamefont {Hazzard}}, \bibinfo {author}
  {\bibfnamefont {A.~M.}\ \bibnamefont {Rey}}, \bibinfo {author} {\bibfnamefont
  {D.~S.}\ \bibnamefont {Jin}}, \ and\ \bibinfo {author} {\bibfnamefont
  {J.}~\bibnamefont {Ye}},\ }\href {http://dx.doi.org/10.1038/nature12483}
  {\bibfield  {journal} {\bibinfo  {journal} {Nature}\ }\textbf {\bibinfo
  {volume} {501}},\ \bibinfo {pages} {521} (\bibinfo {year}
  {2013})}\BibitemShut {NoStop}%
\bibitem [{\citenamefont {Baranov}\ \emph {et~al.}(2012)\citenamefont
  {Baranov}, \citenamefont {Dalmonte}, \citenamefont {Pupillo},\ and\
  \citenamefont {Zoller}}]{Baranov2012}%
  \BibitemOpen
  \bibfield  {author} {\bibinfo {author} {\bibfnamefont {M.~A.}\ \bibnamefont
  {Baranov}}, \bibinfo {author} {\bibfnamefont {M.}~\bibnamefont {Dalmonte}},
  \bibinfo {author} {\bibfnamefont {G.}~\bibnamefont {Pupillo}}, \ and\
  \bibinfo {author} {\bibfnamefont {P.}~\bibnamefont {Zoller}},\ }\href
  {\doibase 10.1021/cr2003568} {\bibfield  {journal} {\bibinfo  {journal}
  {Chem. Rev.}\ }\textbf {\bibinfo {volume} {112}},\ \bibinfo {pages} {5012}
  (\bibinfo {year} {2012})}\BibitemShut {NoStop}%
\bibitem [{\citenamefont {Ospelkaus}\ \emph {et~al.}(2010)\citenamefont
  {Ospelkaus}, \citenamefont {Ni}, \citenamefont {Wang}, \citenamefont
  {de~Miranda}, \citenamefont {Neyenhuis}, \citenamefont {Qu\'em\'ener},
  \citenamefont {Julienne}, \citenamefont {Bohn}, \citenamefont {Jin},\ and\
  \citenamefont {Ye}}]{Ospelkaus2010}%
  \BibitemOpen
  \bibfield  {author} {\bibinfo {author} {\bibfnamefont {S.}~\bibnamefont
  {Ospelkaus}}, \bibinfo {author} {\bibfnamefont {K.-K.}\ \bibnamefont {Ni}},
  \bibinfo {author} {\bibfnamefont {D.}~\bibnamefont {Wang}}, \bibinfo {author}
  {\bibfnamefont {M.~H.~G.}\ \bibnamefont {de~Miranda}}, \bibinfo {author}
  {\bibfnamefont {B.}~\bibnamefont {Neyenhuis}}, \bibinfo {author}
  {\bibfnamefont {G.}~\bibnamefont {Qu\'em\'ener}}, \bibinfo {author}
  {\bibfnamefont {P.~S.}\ \bibnamefont {Julienne}}, \bibinfo {author}
  {\bibfnamefont {J.~L.}\ \bibnamefont {Bohn}}, \bibinfo {author}
  {\bibfnamefont {D.~S.}\ \bibnamefont {Jin}}, \ and\ \bibinfo {author}
  {\bibfnamefont {J.}~\bibnamefont {Ye}},\ }\href {\doibase
  10.1126/science.1184121} {\bibfield  {journal} {\bibinfo  {journal}
  {Science}\ }\textbf {\bibinfo {volume} {327}},\ \bibinfo {pages} {853}
  (\bibinfo {year} {2010})}\BibitemShut {NoStop}%
\bibitem [{\citenamefont {Chin}\ \emph {et~al.}(2009)\citenamefont {Chin},
  \citenamefont {Flambaum},\ and\ \citenamefont {Kozlov}}]{Chin2009}%
  \BibitemOpen
  \bibfield  {author} {\bibinfo {author} {\bibfnamefont {C.}~\bibnamefont
  {Chin}}, \bibinfo {author} {\bibfnamefont {V.~V.}\ \bibnamefont {Flambaum}},
  \ and\ \bibinfo {author} {\bibfnamefont {M.~G.}\ \bibnamefont {Kozlov}},\
  }\href {http://stacks.iop.org/1367-2630/11/i=5/a=055048} {\bibfield
  {journal} {\bibinfo  {journal} {New J. Phys.}\ }\textbf {\bibinfo {volume}
  {11}},\ \bibinfo {pages} {055048} (\bibinfo {year} {2009})}\BibitemShut
  {NoStop}%
\bibitem [{\citenamefont {DeMille}(2002)}]{DeMille2002}%
  \BibitemOpen
  \bibfield  {author} {\bibinfo {author} {\bibfnamefont {D.}~\bibnamefont
  {DeMille}},\ }\href {\doibase 10.1103/PhysRevLett.88.067901} {\bibfield
  {journal} {\bibinfo  {journal} {Phys. Rev. Lett.}\ }\textbf {\bibinfo
  {volume} {88}},\ \bibinfo {pages} {067901} (\bibinfo {year}
  {2002})}\BibitemShut {NoStop}%
\bibitem [{\citenamefont {Andre}\ \emph {et~al.}(2006)\citenamefont {Andre},
  \citenamefont {DeMille}, \citenamefont {Doyle}, \citenamefont {Lukin},
  \citenamefont {Maxwell}, \citenamefont {Rabl}, \citenamefont {Schoelkopf},\
  and\ \citenamefont {Zoller}}]{Andre2006}%
  \BibitemOpen
  \bibfield  {author} {\bibinfo {author} {\bibfnamefont {A.}~\bibnamefont
  {Andre}}, \bibinfo {author} {\bibfnamefont {D.}~\bibnamefont {DeMille}},
  \bibinfo {author} {\bibfnamefont {J.~M.}\ \bibnamefont {Doyle}}, \bibinfo
  {author} {\bibfnamefont {M.~D.}\ \bibnamefont {Lukin}}, \bibinfo {author}
  {\bibfnamefont {S.~E.}\ \bibnamefont {Maxwell}}, \bibinfo {author}
  {\bibfnamefont {P.}~\bibnamefont {Rabl}}, \bibinfo {author} {\bibfnamefont
  {R.~J.}\ \bibnamefont {Schoelkopf}}, \ and\ \bibinfo {author} {\bibfnamefont
  {P.}~\bibnamefont {Zoller}},\ }\href {\doibase
  http://www.nature.com/nphys/journal/v2/n9/suppinfo/nphys386_S1.html}
  {\bibfield  {journal} {\bibinfo  {journal} {Nat. Phys.}\ }\textbf {\bibinfo
  {volume} {2}},\ \bibinfo {pages} {636} (\bibinfo {year} {2006})}\BibitemShut
  {NoStop}%
\bibitem [{\citenamefont {Ni}\ \emph {et~al.}(2008)\citenamefont {Ni},
  \citenamefont {Ospelkaus}, \citenamefont {de~Miranda}, \citenamefont {Pe'er},
  \citenamefont {Neyenhuis}, \citenamefont {Zirbel}, \citenamefont
  {Kotochigova}, \citenamefont {Julienne}, \citenamefont {Jin},\ and\
  \citenamefont {Ye}}]{Ni2008}%
  \BibitemOpen
  \bibfield  {author} {\bibinfo {author} {\bibfnamefont {K.-K.}\ \bibnamefont
  {Ni}}, \bibinfo {author} {\bibfnamefont {S.}~\bibnamefont {Ospelkaus}},
  \bibinfo {author} {\bibfnamefont {M.~H.~G.}\ \bibnamefont {de~Miranda}},
  \bibinfo {author} {\bibfnamefont {A.}~\bibnamefont {Pe'er}}, \bibinfo
  {author} {\bibfnamefont {B.}~\bibnamefont {Neyenhuis}}, \bibinfo {author}
  {\bibfnamefont {J.~J.}\ \bibnamefont {Zirbel}}, \bibinfo {author}
  {\bibfnamefont {S.}~\bibnamefont {Kotochigova}}, \bibinfo {author}
  {\bibfnamefont {P.~S.}\ \bibnamefont {Julienne}}, \bibinfo {author}
  {\bibfnamefont {D.~S.}\ \bibnamefont {Jin}}, \ and\ \bibinfo {author}
  {\bibfnamefont {J.}~\bibnamefont {Ye}},\ }\href {\doibase
  10.1126/science.1163861} {\bibfield  {journal} {\bibinfo  {journal}
  {Science}\ }\textbf {\bibinfo {volume} {322}},\ \bibinfo {pages} {231}
  (\bibinfo {year} {2008})}\BibitemShut {NoStop}%
\bibitem [{\citenamefont {Moses}\ \emph {et~al.}(2015)\citenamefont {Moses},
  \citenamefont {Covey}, \citenamefont {Miecnikowski}, \citenamefont {Yan},
  \citenamefont {Gadway}, \citenamefont {Ye},\ and\ \citenamefont
  {Jin}}]{Moses2015}%
  \BibitemOpen
  \bibfield  {author} {\bibinfo {author} {\bibfnamefont {S.~A.}\ \bibnamefont
  {Moses}}, \bibinfo {author} {\bibfnamefont {J.~P.}\ \bibnamefont {Covey}},
  \bibinfo {author} {\bibfnamefont {M.~T.}\ \bibnamefont {Miecnikowski}},
  \bibinfo {author} {\bibfnamefont {B.}~\bibnamefont {Yan}}, \bibinfo {author}
  {\bibfnamefont {B.}~\bibnamefont {Gadway}}, \bibinfo {author} {\bibfnamefont
  {J.}~\bibnamefont {Ye}}, \ and\ \bibinfo {author} {\bibfnamefont {D.~S.}\
  \bibnamefont {Jin}},\ }\href {\doibase 10.1126/science.aac6400} {\bibfield
  {journal} {\bibinfo  {journal} {Science}\ }\textbf {\bibinfo {volume}
  {350}},\ \bibinfo {pages} {659} (\bibinfo {year} {2015})}\BibitemShut
  {NoStop}%
\bibitem [{\citenamefont {Bochinski}\ \emph {et~al.}(2003)\citenamefont
  {Bochinski}, \citenamefont {Hudson}, \citenamefont {Lewandowski},
  \citenamefont {Meijer},\ and\ \citenamefont {Ye}}]{Bochinski2003}%
  \BibitemOpen
  \bibfield  {author} {\bibinfo {author} {\bibfnamefont {J.~R.}\ \bibnamefont
  {Bochinski}}, \bibinfo {author} {\bibfnamefont {E.~R.}\ \bibnamefont
  {Hudson}}, \bibinfo {author} {\bibfnamefont {H.~J.}\ \bibnamefont
  {Lewandowski}}, \bibinfo {author} {\bibfnamefont {G.}~\bibnamefont {Meijer}},
  \ and\ \bibinfo {author} {\bibfnamefont {J.}~\bibnamefont {Ye}},\ }\href
  {\doibase 10.1103/PhysRevLett.91.243001} {\bibfield  {journal} {\bibinfo
  {journal} {Phys. Rev. Lett.}\ }\textbf {\bibinfo {volume} {91}},\ \bibinfo
  {pages} {243001} (\bibinfo {year} {2003})}\BibitemShut {NoStop}%
\bibitem [{\citenamefont {Narevicius}\ \emph {et~al.}(2008)\citenamefont
  {Narevicius}, \citenamefont {Libson}, \citenamefont {Parthey}, \citenamefont
  {Chavez}, \citenamefont {Narevicius}, \citenamefont {Even},\ and\
  \citenamefont {Raizen}}]{Narevicius2008}%
  \BibitemOpen
  \bibfield  {author} {\bibinfo {author} {\bibfnamefont {E.}~\bibnamefont
  {Narevicius}}, \bibinfo {author} {\bibfnamefont {A.}~\bibnamefont {Libson}},
  \bibinfo {author} {\bibfnamefont {C.~G.}\ \bibnamefont {Parthey}}, \bibinfo
  {author} {\bibfnamefont {I.}~\bibnamefont {Chavez}}, \bibinfo {author}
  {\bibfnamefont {J.}~\bibnamefont {Narevicius}}, \bibinfo {author}
  {\bibfnamefont {U.}~\bibnamefont {Even}}, \ and\ \bibinfo {author}
  {\bibfnamefont {M.~G.}\ \bibnamefont {Raizen}},\ }\href {\doibase
  10.1103/PhysRevLett.100.093003} {\bibfield  {journal} {\bibinfo  {journal}
  {Phys. Rev. Lett.}\ }\textbf {\bibinfo {volume} {100}},\ \bibinfo {pages}
  {093003} (\bibinfo {year} {2008})}\BibitemShut {NoStop}%
\bibitem [{\citenamefont {Fulton}\ \emph {et~al.}(2006)\citenamefont {Fulton},
  \citenamefont {Bishop}, \citenamefont {Shneider},\ and\ \citenamefont
  {Barker}}]{Fulton2006}%
  \BibitemOpen
  \bibfield  {author} {\bibinfo {author} {\bibfnamefont {R.}~\bibnamefont
  {Fulton}}, \bibinfo {author} {\bibfnamefont {A.~I.}\ \bibnamefont {Bishop}},
  \bibinfo {author} {\bibfnamefont {M.~N.}\ \bibnamefont {Shneider}}, \ and\
  \bibinfo {author} {\bibfnamefont {P.~F.}\ \bibnamefont {Barker}},\ }\href
  {http://dx.doi.org/10.1038/nphys339} {\bibfield  {journal} {\bibinfo
  {journal} {Nat. Phys.}\ }\textbf {\bibinfo {volume} {2}},\ \bibinfo {pages}
  {465} (\bibinfo {year} {2006})}\BibitemShut {NoStop}%
\bibitem [{\citenamefont {Liu}\ \emph {et~al.}(2012)\citenamefont {Liu},
  \citenamefont {Yin},\ and\ \citenamefont {Yin}}]{Liu2012}%
  \BibitemOpen
  \bibfield  {author} {\bibinfo {author} {\bibfnamefont {R.-Q.}\ \bibnamefont
  {Liu}}, \bibinfo {author} {\bibfnamefont {Y.-L.}\ \bibnamefont {Yin}}, \ and\
  \bibinfo {author} {\bibfnamefont {J.-P.}\ \bibnamefont {Yin}},\ }\href
  {http://stacks.iop.org/1674-1056/21/i=3/a=033302} {\bibfield  {journal}
  {\bibinfo  {journal} {Chin. Phys. B}\ }\textbf {\bibinfo {volume} {21}},\
  \bibinfo {pages} {033302} (\bibinfo {year} {2012})}\BibitemShut {NoStop}%
\bibitem [{\citenamefont {Di~Rosa}(2004)}]{DiRosa2004}%
  \BibitemOpen
  \bibfield  {author} {\bibinfo {author} {\bibfnamefont {M.~D.}\ \bibnamefont
  {Di~Rosa}},\ }\href {\doibase 10.1140/epjd/e2004-00167-2} {\bibfield
  {journal} {\bibinfo  {journal} {The European Physical Journal D - Atomic,
  Molecular, Optical and Plasma Physics}\ }\textbf {\bibinfo {volume} {31}},\
  \bibinfo {pages} {395} (\bibinfo {year} {2004})}\BibitemShut {NoStop}%
\bibitem [{\citenamefont {Stuhl}\ \emph {et~al.}(2008)\citenamefont {Stuhl},
  \citenamefont {Sawyer}, \citenamefont {Wang},\ and\ \citenamefont
  {Ye}}]{Stuhl2008}%
  \BibitemOpen
  \bibfield  {author} {\bibinfo {author} {\bibfnamefont {B.~K.}\ \bibnamefont
  {Stuhl}}, \bibinfo {author} {\bibfnamefont {B.~C.}\ \bibnamefont {Sawyer}},
  \bibinfo {author} {\bibfnamefont {D.}~\bibnamefont {Wang}}, \ and\ \bibinfo
  {author} {\bibfnamefont {J.}~\bibnamefont {Ye}},\ }\href {\doibase
  10.1103/PhysRevLett.101.243002} {\bibfield  {journal} {\bibinfo  {journal}
  {Phys. Rev. Lett.}\ }\textbf {\bibinfo {volume} {101}},\ \bibinfo {pages}
  {243002} (\bibinfo {year} {2008})}\BibitemShut {NoStop}%
\bibitem [{\citenamefont {Shuman}\ \emph {et~al.}(2010)\citenamefont {Shuman},
  \citenamefont {Barry},\ and\ \citenamefont {DeMille}}]{Shuman2010}%
  \BibitemOpen
  \bibfield  {author} {\bibinfo {author} {\bibfnamefont {E.~S.}\ \bibnamefont
  {Shuman}}, \bibinfo {author} {\bibfnamefont {J.~F.}\ \bibnamefont {Barry}}, \
  and\ \bibinfo {author} {\bibfnamefont {D.}~\bibnamefont {DeMille}},\ }\href
  {\doibase 10.1038/nature09443} {\bibfield  {journal} {\bibinfo  {journal}
  {Nature}\ }\textbf {\bibinfo {volume} {467}},\ \bibinfo {pages} {820}
  (\bibinfo {year} {2010})}\BibitemShut {NoStop}%
\bibitem [{\citenamefont {Barry}\ \emph {et~al.}(2014)\citenamefont {Barry},
  \citenamefont {McCarron}, \citenamefont {Norrgard}, \citenamefont
  {Steinecker},\ and\ \citenamefont {DeMille}}]{Barry2014}%
  \BibitemOpen
  \bibfield  {author} {\bibinfo {author} {\bibfnamefont {J.~F.}\ \bibnamefont
  {Barry}}, \bibinfo {author} {\bibfnamefont {D.~J.}\ \bibnamefont {McCarron}},
  \bibinfo {author} {\bibfnamefont {E.~B.}\ \bibnamefont {Norrgard}}, \bibinfo
  {author} {\bibfnamefont {M.~H.}\ \bibnamefont {Steinecker}}, \ and\ \bibinfo
  {author} {\bibfnamefont {D.}~\bibnamefont {DeMille}},\ }\href@noop {}
  {\bibfield  {journal} {\bibinfo  {journal} {Nature}\ }\textbf {\bibinfo
  {volume} {512}},\ \bibinfo {pages} {286} (\bibinfo {year}
  {2014})}\BibitemShut {NoStop}%
\bibitem [{\citenamefont {Hummon}\ \emph {et~al.}(2013)\citenamefont {Hummon},
  \citenamefont {Yeo}, \citenamefont {Stuhl}, \citenamefont {Collopy},
  \citenamefont {Xia},\ and\ \citenamefont {Ye}}]{Hummon2013}%
  \BibitemOpen
  \bibfield  {author} {\bibinfo {author} {\bibfnamefont {M.~T.}\ \bibnamefont
  {Hummon}}, \bibinfo {author} {\bibfnamefont {M.}~\bibnamefont {Yeo}},
  \bibinfo {author} {\bibfnamefont {B.~K.}\ \bibnamefont {Stuhl}}, \bibinfo
  {author} {\bibfnamefont {A.~L.}\ \bibnamefont {Collopy}}, \bibinfo {author}
  {\bibfnamefont {Y.}~\bibnamefont {Xia}}, \ and\ \bibinfo {author}
  {\bibfnamefont {J.}~\bibnamefont {Ye}},\ }\href {\doibase
  10.1103/PhysRevLett.110.143001} {\bibfield  {journal} {\bibinfo  {journal}
  {Phys. Rev. Lett.}\ }\textbf {\bibinfo {volume} {110}},\ \bibinfo {pages}
  {143001} (\bibinfo {year} {2013})}\BibitemShut {NoStop}%
\bibitem [{\citenamefont {Yeo}\ \emph {et~al.}(2015)\citenamefont {Yeo},
  \citenamefont {Hummon}, \citenamefont {Collopy}, \citenamefont {Yan},
  \citenamefont {Hemmerling}, \citenamefont {Chae}, \citenamefont {Doyle},\
  and\ \citenamefont {Ye}}]{Yeo2015}%
  \BibitemOpen
  \bibfield  {author} {\bibinfo {author} {\bibfnamefont {M.}~\bibnamefont
  {Yeo}}, \bibinfo {author} {\bibfnamefont {M.~T.}\ \bibnamefont {Hummon}},
  \bibinfo {author} {\bibfnamefont {A.~L.}\ \bibnamefont {Collopy}}, \bibinfo
  {author} {\bibfnamefont {B.}~\bibnamefont {Yan}}, \bibinfo {author}
  {\bibfnamefont {B.}~\bibnamefont {Hemmerling}}, \bibinfo {author}
  {\bibfnamefont {E.}~\bibnamefont {Chae}}, \bibinfo {author} {\bibfnamefont
  {J.~M.}\ \bibnamefont {Doyle}}, \ and\ \bibinfo {author} {\bibfnamefont
  {J.}~\bibnamefont {Ye}},\ }\href {\doibase 10.1103/PhysRevLett.114.223003}
  {\bibfield  {journal} {\bibinfo  {journal} {Phys. Rev. Lett.}\ }\textbf
  {\bibinfo {volume} {114}},\ \bibinfo {pages} {223003} (\bibinfo {year}
  {2015})}\BibitemShut {NoStop}%
\bibitem [{\citenamefont {Zhelyazkova}\ \emph {et~al.}(2014)\citenamefont
  {Zhelyazkova}, \citenamefont {Cournol}, \citenamefont {Wall}, \citenamefont
  {Matsushima}, \citenamefont {Hudson}, \citenamefont {Hinds}, \citenamefont
  {Tarbutt},\ and\ \citenamefont {Sauer}}]{Zhelyazkova2014}%
  \BibitemOpen
  \bibfield  {author} {\bibinfo {author} {\bibfnamefont {V.}~\bibnamefont
  {Zhelyazkova}}, \bibinfo {author} {\bibfnamefont {A.}~\bibnamefont
  {Cournol}}, \bibinfo {author} {\bibfnamefont {T.~E.}\ \bibnamefont {Wall}},
  \bibinfo {author} {\bibfnamefont {A.}~\bibnamefont {Matsushima}}, \bibinfo
  {author} {\bibfnamefont {J.~J.}\ \bibnamefont {Hudson}}, \bibinfo {author}
  {\bibfnamefont {E.~A.}\ \bibnamefont {Hinds}}, \bibinfo {author}
  {\bibfnamefont {M.~R.}\ \bibnamefont {Tarbutt}}, \ and\ \bibinfo {author}
  {\bibfnamefont {B.~E.}\ \bibnamefont {Sauer}},\ }\href {\doibase
  10.1103/PhysRevA.89.053416} {\bibfield  {journal} {\bibinfo  {journal} {Phys.
  Rev. A}\ }\textbf {\bibinfo {volume} {89}},\ \bibinfo {pages} {053416}
  (\bibinfo {year} {2014})}\BibitemShut {NoStop}%
\bibitem [{\citenamefont {Hemmerling}\ \emph {et~al.}(2016)\citenamefont
  {Hemmerling}, \citenamefont {Chae}, \citenamefont {Ravi}, \citenamefont
  {Anderegg}, \citenamefont {Drayna}, \citenamefont {Hutzler}, \citenamefont
  {Collopy}, \citenamefont {Ye}, \citenamefont {Ketterle},\ and\ \citenamefont
  {Doyle}}]{Hemmerling2016}%
  \BibitemOpen
  \bibfield  {author} {\bibinfo {author} {\bibfnamefont {B.}~\bibnamefont
  {Hemmerling}}, \bibinfo {author} {\bibfnamefont {E.}~\bibnamefont {Chae}},
  \bibinfo {author} {\bibfnamefont {A.}~\bibnamefont {Ravi}}, \bibinfo {author}
  {\bibfnamefont {L.}~\bibnamefont {Anderegg}}, \bibinfo {author}
  {\bibfnamefont {G.~K.}\ \bibnamefont {Drayna}}, \bibinfo {author}
  {\bibfnamefont {N.~R.}\ \bibnamefont {Hutzler}}, \bibinfo {author}
  {\bibfnamefont {A.~L.}\ \bibnamefont {Collopy}}, \bibinfo {author}
  {\bibfnamefont {J.}~\bibnamefont {Ye}}, \bibinfo {author} {\bibfnamefont
  {W.}~\bibnamefont {Ketterle}}, \ and\ \bibinfo {author} {\bibfnamefont
  {J.~M.}\ \bibnamefont {Doyle}},\ }\href {http://arxiv.org/abs/1603.02787}
  {\bibfield  {journal} {\bibinfo  {journal} {arXiv}\ }\textbf {\bibinfo
  {volume} {1603}},\ \bibinfo {pages} {02787} (\bibinfo {year}
  {2016})}\BibitemShut {NoStop}%
\bibitem [{\citenamefont {Smallman}\ \emph {et~al.}(2014)\citenamefont
  {Smallman}, \citenamefont {Wang}, \citenamefont {Steimle}, \citenamefont
  {Tarbutt},\ and\ \citenamefont {Hinds}}]{Smallman2014}%
  \BibitemOpen
  \bibfield  {author} {\bibinfo {author} {\bibfnamefont {I.}~\bibnamefont
  {Smallman}}, \bibinfo {author} {\bibfnamefont {F.}~\bibnamefont {Wang}},
  \bibinfo {author} {\bibfnamefont {T.}~\bibnamefont {Steimle}}, \bibinfo
  {author} {\bibfnamefont {M.}~\bibnamefont {Tarbutt}}, \ and\ \bibinfo
  {author} {\bibfnamefont {E.}~\bibnamefont {Hinds}},\ }\href {\doibase
  http://dx.doi.org/10.1016/j.jms.2014.02.006} {\bibfield  {journal} {\bibinfo
  {journal} {J. Mol. Spectrosc.}\ }\textbf {\bibinfo {volume} {300}},\ \bibinfo
  {pages} {3 } (\bibinfo {year} {2014})}\BibitemShut {NoStop}%
\bibitem [{\citenamefont {Xu}\ \emph {et~al.}(2016)\citenamefont {Xu},
  \citenamefont {Yin}, \citenamefont {Wei}, \citenamefont {Xia},\ and\
  \citenamefont {Yin}}]{Xu2016}%
  \BibitemOpen
  \bibfield  {author} {\bibinfo {author} {\bibfnamefont {L.}~\bibnamefont
  {Xu}}, \bibinfo {author} {\bibfnamefont {Y.}~\bibnamefont {Yin}}, \bibinfo
  {author} {\bibfnamefont {B.}~\bibnamefont {Wei}}, \bibinfo {author}
  {\bibfnamefont {Y.}~\bibnamefont {Xia}}, \ and\ \bibinfo {author}
  {\bibfnamefont {J.}~\bibnamefont {Yin}},\ }\href {\doibase
  10.1103/PhysRevA.93.013408} {\bibfield  {journal} {\bibinfo  {journal} {Phys.
  Rev. A}\ }\textbf {\bibinfo {volume} {93}},\ \bibinfo {pages} {013408}
  (\bibinfo {year} {2016})}\BibitemShut {NoStop}%
\bibitem [{\citenamefont {Hendricks}\ \emph {et~al.}(2014)\citenamefont
  {Hendricks}, \citenamefont {Holland}, \citenamefont {Truppe}, \citenamefont
  {Sauer},\ and\ \citenamefont {Tarbutt}}]{Hendricks2014}%
  \BibitemOpen
  \bibfield  {author} {\bibinfo {author} {\bibfnamefont {R.~J.}\ \bibnamefont
  {Hendricks}}, \bibinfo {author} {\bibfnamefont {D.~A.}\ \bibnamefont
  {Holland}}, \bibinfo {author} {\bibfnamefont {S.}~\bibnamefont {Truppe}},
  \bibinfo {author} {\bibfnamefont {B.~E.}\ \bibnamefont {Sauer}}, \ and\
  \bibinfo {author} {\bibfnamefont {M.~R.}\ \bibnamefont {Tarbutt}},\ }\href
  {\doibase 10.3389/fphy.2014.00051} {\bibfield  {journal} {\bibinfo  {journal}
  {Front. Phys.}\ }\textbf {\bibinfo {volume} {2}},\ \bibinfo {pages} {51}
  (\bibinfo {year} {2014})}\BibitemShut {NoStop}%
\bibitem [{\citenamefont {Isaev}\ \emph {et~al.}(2010)\citenamefont {Isaev},
  \citenamefont {Hoekstra},\ and\ \citenamefont {Berger}}]{Isaev2010}%
  \BibitemOpen
  \bibfield  {author} {\bibinfo {author} {\bibfnamefont {T.~A.}\ \bibnamefont
  {Isaev}}, \bibinfo {author} {\bibfnamefont {S.}~\bibnamefont {Hoekstra}}, \
  and\ \bibinfo {author} {\bibfnamefont {R.}~\bibnamefont {Berger}},\ }\href
  {\doibase {10.1103/PhysRevA.82.052521}} {\bibfield  {journal} {\bibinfo
  {journal} {Phys. Rev. A}\ }\textbf {\bibinfo {volume} {{82}}},\ \bibinfo
  {pages} {{052521}} (\bibinfo {year} {2010})}\BibitemShut {NoStop}%
\bibitem [{\citenamefont {Hunter}\ \emph {et~al.}(2012)\citenamefont {Hunter},
  \citenamefont {Peck}, \citenamefont {Greenspon}, \citenamefont {Alam},\ and\
  \citenamefont {DeMille}}]{Hunter2012}%
  \BibitemOpen
  \bibfield  {author} {\bibinfo {author} {\bibfnamefont {L.~R.}\ \bibnamefont
  {Hunter}}, \bibinfo {author} {\bibfnamefont {S.~K.}\ \bibnamefont {Peck}},
  \bibinfo {author} {\bibfnamefont {A.~S.}\ \bibnamefont {Greenspon}}, \bibinfo
  {author} {\bibfnamefont {S.~S.}\ \bibnamefont {Alam}}, \ and\ \bibinfo
  {author} {\bibfnamefont {D.}~\bibnamefont {DeMille}},\ }\href {\doibase
  10.1103/PhysRevA.85.012511} {\bibfield  {journal} {\bibinfo  {journal} {Phys.
  Rev. A}\ }\textbf {\bibinfo {volume} {85}},\ \bibinfo {pages} {012511}
  (\bibinfo {year} {2012})}\BibitemShut {NoStop}%
\bibitem [{\citenamefont {Tarallo}\ \emph {et~al.}(2016)\citenamefont
  {Tarallo}, \citenamefont {Iwata},\ and\ \citenamefont
  {Zelevinsky}}]{Tarallo2016}%
  \BibitemOpen
  \bibfield  {author} {\bibinfo {author} {\bibfnamefont {M.~G.}\ \bibnamefont
  {Tarallo}}, \bibinfo {author} {\bibfnamefont {G.~Z.}\ \bibnamefont {Iwata}},
  \ and\ \bibinfo {author} {\bibfnamefont {T.}~\bibnamefont {Zelevinsky}},\
  }\href {\doibase 10.1103/PhysRevA.93.032509} {\bibfield  {journal} {\bibinfo
  {journal} {Phys. Rve. A}\ }\textbf {\bibinfo {volume} {93}},\ \bibinfo
  {pages} {032509} (\bibinfo {year} {2016})}\BibitemShut {NoStop}%
\bibitem [{\citenamefont {Norrgard}\ \emph {et~al.}(2016)\citenamefont
  {Norrgard}, \citenamefont {McCarron}, \citenamefont {Steinecker},
  \citenamefont {Tarbutt},\ and\ \citenamefont {DeMille}}]{Norrgard2016}%
  \BibitemOpen
  \bibfield  {author} {\bibinfo {author} {\bibfnamefont {E.~B.}\ \bibnamefont
  {Norrgard}}, \bibinfo {author} {\bibfnamefont {D.~J.}\ \bibnamefont
  {McCarron}}, \bibinfo {author} {\bibfnamefont {M.~H.}\ \bibnamefont
  {Steinecker}}, \bibinfo {author} {\bibfnamefont {M.~R.}\ \bibnamefont
  {Tarbutt}}, \ and\ \bibinfo {author} {\bibfnamefont {D.}~\bibnamefont
  {DeMille}},\ }\href {\doibase 10.1103/PhysRevLett.116.063004} {\bibfield
  {journal} {\bibinfo  {journal} {Phys. Rev. Lett.}\ }\textbf {\bibinfo
  {volume} {116}},\ \bibinfo {pages} {063004} (\bibinfo {year}
  {2016})}\BibitemShut {NoStop}%
\bibitem [{\citenamefont {DeMille}\ \emph {et~al.}(2008)\citenamefont
  {DeMille}, \citenamefont {Cahn}, \citenamefont {Murphree}, \citenamefont
  {Rahmlow},\ and\ \citenamefont {Kozlov}}]{DeMille2008}%
  \BibitemOpen
  \bibfield  {author} {\bibinfo {author} {\bibfnamefont {D.}~\bibnamefont
  {DeMille}}, \bibinfo {author} {\bibfnamefont {S.~B.}\ \bibnamefont {Cahn}},
  \bibinfo {author} {\bibfnamefont {D.}~\bibnamefont {Murphree}}, \bibinfo
  {author} {\bibfnamefont {D.~A.}\ \bibnamefont {Rahmlow}}, \ and\ \bibinfo
  {author} {\bibfnamefont {M.~G.}\ \bibnamefont {Kozlov}},\ }\href {\doibase
  10.1103/PhysRevLett.100.023003} {\bibfield  {journal} {\bibinfo  {journal}
  {Phys. Rev. Lett.}\ }\textbf {\bibinfo {volume} {100}},\ \bibinfo {pages}
  {023003} (\bibinfo {year} {2008})}\BibitemShut {NoStop}%
\bibitem [{\citenamefont {Berg}\ \emph {et~al.}(1998)\citenamefont {Berg},
  \citenamefont {Gador}, \citenamefont {Husain}, \citenamefont {Ludwigs},\ and\
  \citenamefont {Royen}}]{Berg1998}%
  \BibitemOpen
  \bibfield  {author} {\bibinfo {author} {\bibfnamefont {L.-E.}\ \bibnamefont
  {Berg}}, \bibinfo {author} {\bibfnamefont {N.}~\bibnamefont {Gador}},
  \bibinfo {author} {\bibfnamefont {D.}~\bibnamefont {Husain}}, \bibinfo
  {author} {\bibfnamefont {H.}~\bibnamefont {Ludwigs}}, \ and\ \bibinfo
  {author} {\bibfnamefont {P.}~\bibnamefont {Royen}},\ }\href {\doibase
  http://dx.doi.org/10.1016/S0009-2614(98)00149-3} {\bibfield  {journal}
  {\bibinfo  {journal} {Chem. Phys. Lett.}\ }\textbf {\bibinfo {volume}
  {287}},\ \bibinfo {pages} {89 } (\bibinfo {year} {1998})}\BibitemShut
  {NoStop}%
\bibitem [{\citenamefont {Bernard}\ \emph {et~al.}(1992)\citenamefont
  {Bernard}, \citenamefont {Effantin}, \citenamefont {Andrianavalona},
  \citenamefont {Vergès},\ and\ \citenamefont {Barrow}}]{Bernard1992}%
  \BibitemOpen
  \bibfield  {author} {\bibinfo {author} {\bibfnamefont {A.}~\bibnamefont
  {Bernard}}, \bibinfo {author} {\bibfnamefont {C.}~\bibnamefont {Effantin}},
  \bibinfo {author} {\bibfnamefont {E.}~\bibnamefont {Andrianavalona}},
  \bibinfo {author} {\bibfnamefont {J.}~\bibnamefont {Vergès}}, \ and\
  \bibinfo {author} {\bibfnamefont {R.}~\bibnamefont {Barrow}},\ }\href
  {\doibase http://dx.doi.org/10.1016/0022-2852(92)90127-A} {\bibfield
  {journal} {\bibinfo  {journal} {J. Mol. Spectrosc.}\ }\textbf {\bibinfo
  {volume} {152}},\ \bibinfo {pages} {174 } (\bibinfo {year}
  {1992})}\BibitemShut {NoStop}%
\bibitem [{\citenamefont {Rees}(1947)}]{Rees1947}%
  \BibitemOpen
  \bibfield  {author} {\bibinfo {author} {\bibfnamefont {A.~L.~G.}\
  \bibnamefont {Rees}},\ }\href {http://stacks.iop.org/0959-5309/59/i=6/a=310}
  {\bibfield  {journal} {\bibinfo  {journal} {Proceedings of the Physical
  Society}\ }\textbf {\bibinfo {volume} {59}},\ \bibinfo {pages} {998}
  (\bibinfo {year} {1947})}\BibitemShut {NoStop}%
\bibitem [{\citenamefont {Chen}\ \emph {et~al.}(2014)\citenamefont {Chen},
  \citenamefont {Zhu}, \citenamefont {Li}, \citenamefont {Qian},\ and\
  \citenamefont {Wang}}]{Chen2014}%
  \BibitemOpen
  \bibfield  {author} {\bibinfo {author} {\bibfnamefont {T.}~\bibnamefont
  {Chen}}, \bibinfo {author} {\bibfnamefont {S.}~\bibnamefont {Zhu}}, \bibinfo
  {author} {\bibfnamefont {X.}~\bibnamefont {Li}}, \bibinfo {author}
  {\bibfnamefont {J.}~\bibnamefont {Qian}}, \ and\ \bibinfo {author}
  {\bibfnamefont {Y.}~\bibnamefont {Wang}},\ }\href {\doibase
  10.1103/PhysRevA.89.063402} {\bibfield  {journal} {\bibinfo  {journal} {Phys.
  Rev. A}\ }\textbf {\bibinfo {volume} {89}},\ \bibinfo {pages} {063402}
  (\bibinfo {year} {2014})}\BibitemShut {NoStop}%
\bibitem [{\citenamefont {Steimle}\ \emph {et~al.}(2011)\citenamefont
  {Steimle}, \citenamefont {Frey}, \citenamefont {Le}, \citenamefont {DeMille},
  \citenamefont {Rahmlow},\ and\ \citenamefont {Linton}}]{Steimle2011}%
  \BibitemOpen
  \bibfield  {author} {\bibinfo {author} {\bibfnamefont {T.~C.}\ \bibnamefont
  {Steimle}}, \bibinfo {author} {\bibfnamefont {S.}~\bibnamefont {Frey}},
  \bibinfo {author} {\bibfnamefont {A.}~\bibnamefont {Le}}, \bibinfo {author}
  {\bibfnamefont {D.}~\bibnamefont {DeMille}}, \bibinfo {author} {\bibfnamefont
  {D.~A.}\ \bibnamefont {Rahmlow}}, \ and\ \bibinfo {author} {\bibfnamefont
  {C.}~\bibnamefont {Linton}},\ }\href {\doibase 10.1103/PhysRevA.84.012508}
  {\bibfield  {journal} {\bibinfo  {journal} {Phys. Rev. A}\ }\textbf {\bibinfo
  {volume} {84}},\ \bibinfo {pages} {012508} (\bibinfo {year}
  {2011})}\BibitemShut {NoStop}%
\bibitem [{\citenamefont {Barrow}\ \emph {et~al.}(1988)\citenamefont {Barrow},
  \citenamefont {Bernard}, \citenamefont {Effantin}, \citenamefont {D'Incan},
  \citenamefont {Fabre}, \citenamefont {Hachimi}, \citenamefont {Stringat},\
  and\ \citenamefont {Vergès}}]{Barrow1988}%
  \BibitemOpen
  \bibfield  {author} {\bibinfo {author} {\bibfnamefont {R.}~\bibnamefont
  {Barrow}}, \bibinfo {author} {\bibfnamefont {A.}~\bibnamefont {Bernard}},
  \bibinfo {author} {\bibfnamefont {C.}~\bibnamefont {Effantin}}, \bibinfo
  {author} {\bibfnamefont {J.}~\bibnamefont {D'Incan}}, \bibinfo {author}
  {\bibfnamefont {G.}~\bibnamefont {Fabre}}, \bibinfo {author} {\bibfnamefont
  {A.~E.}\ \bibnamefont {Hachimi}}, \bibinfo {author} {\bibfnamefont
  {R.}~\bibnamefont {Stringat}}, \ and\ \bibinfo {author} {\bibfnamefont
  {J.}~\bibnamefont {Vergès}},\ }\href {\doibase
  http://dx.doi.org/10.1016/0009-2614(88)80263-X} {\bibfield  {journal}
  {\bibinfo  {journal} {Chem. Phys. Lett.}\ }\textbf {\bibinfo {volume}
  {147}},\ \bibinfo {pages} {535 } (\bibinfo {year} {1988})}\BibitemShut
  {NoStop}%
\bibitem [{\citenamefont {Mulliken}\ and\ \citenamefont
  {Christy}(1931)}]{Mulliken1931}%
  \BibitemOpen
  \bibfield  {author} {\bibinfo {author} {\bibfnamefont {R.~S.}\ \bibnamefont
  {Mulliken}}\ and\ \bibinfo {author} {\bibfnamefont {A.}~\bibnamefont
  {Christy}},\ }\href {\doibase 10.1103/PhysRev.38.87} {\bibfield  {journal}
  {\bibinfo  {journal} {Phys. Rev.}\ }\textbf {\bibinfo {volume} {38}},\
  \bibinfo {pages} {87} (\bibinfo {year} {1931})}\BibitemShut {NoStop}%
\bibitem [{\citenamefont {Guo}\ \emph {et~al.}(1995)\citenamefont {Guo},
  \citenamefont {Zhang},\ and\ \citenamefont {Bernath}}]{Guo1995}%
  \BibitemOpen
  \bibfield  {author} {\bibinfo {author} {\bibfnamefont {B.}~\bibnamefont
  {Guo}}, \bibinfo {author} {\bibfnamefont {K.}~\bibnamefont {Zhang}}, \ and\
  \bibinfo {author} {\bibfnamefont {P.}~\bibnamefont {Bernath}},\ }\href
  {\doibase http://dx.doi.org/10.1006/jmsp.1995.1056} {\bibfield  {journal}
  {\bibinfo  {journal} {J. Mol. Spectrosc.}\ }\textbf {\bibinfo {volume}
  {170}},\ \bibinfo {pages} {59 } (\bibinfo {year} {1995})}\BibitemShut
  {NoStop}%
\bibitem [{\citenamefont {Ryzlewicz}\ and\ \citenamefont
  {Törring}(1980)}]{Ryzlewicz1980}%
  \BibitemOpen
  \bibfield  {author} {\bibinfo {author} {\bibfnamefont {C.}~\bibnamefont
  {Ryzlewicz}}\ and\ \bibinfo {author} {\bibfnamefont {T.}~\bibnamefont
  {Törring}},\ }\href {\doibase
  http://dx.doi.org/10.1016/0301-0104(80)80107-8} {\bibfield  {journal}
  {\bibinfo  {journal} {Chem. Phys.}\ }\textbf {\bibinfo {volume} {51}},\
  \bibinfo {pages} {329 } (\bibinfo {year} {1980})}\BibitemShut {NoStop}%
\bibitem [{\citenamefont {Ernst}\ \emph {et~al.}(1986)\citenamefont {Ernst},
  \citenamefont {Kändler},\ and\ \citenamefont {Törring}}]{Ernst1986}%
  \BibitemOpen
  \bibfield  {author} {\bibinfo {author} {\bibfnamefont {W.~E.}\ \bibnamefont
  {Ernst}}, \bibinfo {author} {\bibfnamefont {J.}~\bibnamefont {Kändler}}, \
  and\ \bibinfo {author} {\bibfnamefont {T.}~\bibnamefont {Törring}},\ }\href
  {\doibase http://dx.doi.org/10.1063/1.449961} {\bibfield  {journal} {\bibinfo
   {journal} {The Journal of Chemical Physics}\ }\textbf {\bibinfo {volume}
  {84}},\ \bibinfo {pages} {4769} (\bibinfo {year} {1986})}\BibitemShut
  {NoStop}%
\bibitem [{\citenamefont {Wall}\ \emph {et~al.}(2008)\citenamefont {Wall},
  \citenamefont {Kanem}, \citenamefont {Hudson}, \citenamefont {Sauer},
  \citenamefont {Cho}, \citenamefont {Boshier}, \citenamefont {Hinds},\ and\
  \citenamefont {Tarbutt}}]{Wall2008}%
  \BibitemOpen
  \bibfield  {author} {\bibinfo {author} {\bibfnamefont {T.~E.}\ \bibnamefont
  {Wall}}, \bibinfo {author} {\bibfnamefont {J.~F.}\ \bibnamefont {Kanem}},
  \bibinfo {author} {\bibfnamefont {J.~J.}\ \bibnamefont {Hudson}}, \bibinfo
  {author} {\bibfnamefont {B.~E.}\ \bibnamefont {Sauer}}, \bibinfo {author}
  {\bibfnamefont {D.}~\bibnamefont {Cho}}, \bibinfo {author} {\bibfnamefont
  {M.~G.}\ \bibnamefont {Boshier}}, \bibinfo {author} {\bibfnamefont {E.~A.}\
  \bibnamefont {Hinds}}, \ and\ \bibinfo {author} {\bibfnamefont {M.~R.}\
  \bibnamefont {Tarbutt}},\ }\href {\doibase 10.1103/PhysRevA.78.062509}
  {\bibfield  {journal} {\bibinfo  {journal} {Phys. Rev. A}\ }\textbf {\bibinfo
  {volume} {78}},\ \bibinfo {pages} {062509} (\bibinfo {year}
  {2008})}\BibitemShut {NoStop}%
\bibitem [{\citenamefont {Tarbutt}(2015)}]{Tarbutt2015}%
  \BibitemOpen
  \bibfield  {author} {\bibinfo {author} {\bibfnamefont {M.~R.}\ \bibnamefont
  {Tarbutt}},\ }\href {http://stacks.iop.org/1367-2630/17/i=1/a=015007}
  {\bibfield  {journal} {\bibinfo  {journal} {New J. Phys.}\ }\textbf {\bibinfo
  {volume} {17}},\ \bibinfo {pages} {015007} (\bibinfo {year}
  {2015})}\BibitemShut {NoStop}%
\bibitem [{\citenamefont {Brown}\ and\ \citenamefont
  {Carrington}(2003)}]{Brown2003}%
  \BibitemOpen
  \bibfield  {author} {\bibinfo {author} {\bibfnamefont {J.}~\bibnamefont
  {Brown}}\ and\ \bibinfo {author} {\bibfnamefont {A.}~\bibnamefont
  {Carrington}},\ }\href@noop {} {\emph {\bibinfo {title} {Rotational
  Spectroscopy of Diatomic Molecules}}},\ edited by\ \bibinfo {editor}
  {\bibfnamefont {A.~H.~Z.}\ \bibnamefont {Richard J.~Saykally}}\ and\ \bibinfo
  {editor} {\bibfnamefont {D.~A.}\ \bibnamefont {King}}\ (\bibinfo  {publisher}
  {Cambridge University Press},\ \bibinfo {year} {2003})\BibitemShut {NoStop}%
\bibitem [{\citenamefont {Steimle}\ \emph {et~al.}(1978)\citenamefont
  {Steimle}, \citenamefont {Domaille},\ and\ \citenamefont
  {Harris}}]{Steimle1978}%
  \BibitemOpen
  \bibfield  {author} {\bibinfo {author} {\bibfnamefont {T.~C.}\ \bibnamefont
  {Steimle}}, \bibinfo {author} {\bibfnamefont {P.~J.}\ \bibnamefont
  {Domaille}}, \ and\ \bibinfo {author} {\bibfnamefont {D.~O.}\ \bibnamefont
  {Harris}},\ }\href {\doibase http://dx.doi.org/10.1016/0022-2852(78)90110-8}
  {\bibfield  {journal} {\bibinfo  {journal} {J. Mol. Spectrosc.}\ }\textbf
  {\bibinfo {volume} {73}},\ \bibinfo {pages} {441 } (\bibinfo {year}
  {1978})}\BibitemShut {NoStop}%
\bibitem [{\citenamefont {Nakagawa}\ \emph {et~al.}(1978)\citenamefont
  {Nakagawa}, \citenamefont {Domaille}, \citenamefont {Steimle},\ and\
  \citenamefont {Harris}}]{Nakagawa1978}%
  \BibitemOpen
  \bibfield  {author} {\bibinfo {author} {\bibfnamefont {J.}~\bibnamefont
  {Nakagawa}}, \bibinfo {author} {\bibfnamefont {P.~J.}\ \bibnamefont
  {Domaille}}, \bibinfo {author} {\bibfnamefont {T.~C.}\ \bibnamefont
  {Steimle}}, \ and\ \bibinfo {author} {\bibfnamefont {D.~O.}\ \bibnamefont
  {Harris}},\ }\href {\doibase http://dx.doi.org/10.1016/0022-2852(78)90175-3}
  {\bibfield  {journal} {\bibinfo  {journal} {J. Mol. Spectrosc.}\ }\textbf
  {\bibinfo {volume} {70}},\ \bibinfo {pages} {374 } (\bibinfo {year}
  {1978})}\BibitemShut {NoStop}%
\bibitem [{\citenamefont {Collopy}\ \emph {et~al.}(2015)\citenamefont
  {Collopy}, \citenamefont {Hummon}, \citenamefont {Yeo}, \citenamefont {Yan},\
  and\ \citenamefont {Ye}}]{Collopy2015}%
  \BibitemOpen
  \bibfield  {author} {\bibinfo {author} {\bibfnamefont {A.~L.}\ \bibnamefont
  {Collopy}}, \bibinfo {author} {\bibfnamefont {M.~T.}\ \bibnamefont {Hummon}},
  \bibinfo {author} {\bibfnamefont {M.}~\bibnamefont {Yeo}}, \bibinfo {author}
  {\bibfnamefont {B.}~\bibnamefont {Yan}}, \ and\ \bibinfo {author}
  {\bibfnamefont {J.}~\bibnamefont {Ye}},\ }\href
  {http://stacks.iop.org/1367-2630/17/i=5/a=055008} {\bibfield  {journal}
  {\bibinfo  {journal} {New J. Phys.}\ }\textbf {\bibinfo {volume} {17}},\
  \bibinfo {pages} {055008} (\bibinfo {year} {2015})}\BibitemShut {NoStop}%
\bibitem [{\citenamefont {Chalek}\ and\ \citenamefont
  {Gole}(1976)}]{Chalek1976}%
  \BibitemOpen
  \bibfield  {author} {\bibinfo {author} {\bibfnamefont {C.~L.}\ \bibnamefont
  {Chalek}}\ and\ \bibinfo {author} {\bibfnamefont {J.~L.}\ \bibnamefont
  {Gole}},\ }\href {\doibase http://dx.doi.org/10.1063/1.433434} {\bibfield
  {journal} {\bibinfo  {journal} {The Journal of Chemical Physics}\ }\textbf
  {\bibinfo {volume} {65}},\ \bibinfo {pages} {2845} (\bibinfo {year}
  {1976})}\BibitemShut {NoStop}%
\end{thebibliography}%

\end{document}